\documentclass[manuscript]{aastex}

\slugcomment{Not to appear in Nonlearned J., 45.}

\shorttitle{Loop-loop interaction}
\shortauthors{Kumar et al.}

\begin{document}


\title{EVIDENCE OF SOLAR FLARE TRIGGERING DUE TO LOOP-LOOP INTERACTION
       CAUSED BY FOOTPOINT SHEAR-MOTION}
\author{PANKAJ KUMAR$^{1}$}
\affil{Aryabhatta Research Institute of Observational Sciences (ARIES),
       Nainital-263129, India}
\email{pkumar@aries.res.in}

\author{A. K. SRIVASTAVA$^{1,4}$}
\affil{Aryabhatta Research Institute of Observational Sciences (ARIES),
       Nainital-263129, India}
%\email{aks@aries.res.in}

\author{B. V. SOMOV$^{2}$}
\affil{Astronomical Institute, Moscow State University,
       Universitetskij Prospekt 13, Moscow 119992, Russia}
%\email{somov@sai.msu.ru}

\author{P. K. MANOHARAN$^{3}$}
\affil{Radio Astronomy Centre, NCRA, Tata Institute of Fundamental Research,
       Udhagamandalam (Ooty) 643 001, India}
%\email{mano@ncra.tifr.res.in}

\author{R. ERD\'ELYI$^{4}$}
\affil{Solar Physics and Space Plasma Research Centre (SP$^{2}$RC), Department of Applied Mathematics, The University of Sheffield S3 7RH, U.K.}

\author{WAHAB UDDIN$^{1}$}
\affil{Aryabhatta Research Institute of Observational Sciences (ARIES),
       Nainital-263129, India}


%\altaffiltext{1}{Visiting Astronomer, Cerro Tololo Inter-American Observatory.
%CTIO is operated by AURA, Inc.\ under contract to the National Science
%\altaffiltext{2}{Society of Fellows, Harvard University.}
%\altaffiltext{3}{present address: Center for Astrophysics,
%    60 Garden Street, Cambridge, MA 02138}
%\altaffiltext{4}{Visiting Programmer, Space Telescope Science Institute}
%\altaffiltext{5}{Patron, Alonso's Bar and Grill}

\begin{abstract}
We analyze multi-wavelength data of a M7.9/1N class solar flare which occurred
on 27 April, 2006 from AR NOAA 10875.
GOES soft X-ray images provide the most likely signature of two interacting loops
and their reconnection, which triggers the solar flare.
TRACE 195 \AA \ images also reveal the loop-loop interaction and the formation of
`X' points with converging motion ($\sim$30 km s$^{-1}$) at the reconnection
site in-between this interacting loop system. This provides the evidence of progressive
reconnection and flare maximization at the interaction site in the active region.
The absence of type III radio burst during this time period indicates no opening of 
magnetic field lines during the flare energy release, which implies
only the change of field lines connectivity/orientation during the loop-loop interaction 
and reconnection process.
The Ondrejov dynamic radio spectrum shows an intense decimetric (DCIM) radio burst (2.5--4.5 GHz, duration$\sim$3 min) during
flare initiation, which reveals the signature of particle acceleration from
the reconnection site during loop-loop interaction.
The double peak structures at 4.9 and 8.8 GHz provide the most likely
confirmatory signature of the loop-loop interaction at the flare site in the active region.
RHESSI hard X-ray images also show the loop-top and
footpoint sources of the corresponding two loop system and their coalescence
during the flare maximum, which act like the current carrying flux-tubes with 
resultant opposite magnetic fields and the net force of attraction.
%The two-current loops attract if the currents in the both loops
%is in the same direction or having magnetic field in opposite direction.
%The shear motion/rotation of the foot-points of loop system
%may also play a crucial role in the interaction and flare triggering.
We also suggest that the shear motion/rotation of the footpoint of the smaller
loop, which is anchored in the opposite polarity spot, may be responsible 
for the flare energy buildup and then its release due to the loop-loop interaction.
\end{abstract}

\footnote [1] {Corresponding Author : Dr. A.K. Srivastava (aks@aries.res.in)}
\section{INTRODUCTION}

% Solar flares are the sudden explosions in the solar atmosphere during which
% the magnetic energy stored in the twisted and sheared magnetic fields is
% released in the form of thermal energy to heat-up the ambient plasma and
% bulk particle acceleration.
%%%%%%  Definition is not complited. Maybe, the following is better ??? %%%%%%
% ? %
Solar flare is a sudden explosion in the solar
atmosphere during which the magnetic energy (stored in the twisted and
sheared magnetic fields as well as in the current layers between
interacting fields) is released in the form of kinetic energy of rapidly
moving plasma, accelerated particles and thermal energy to heat-up
the ambient plasma.  This primary release of energy takes place in the corona and is
accompanied by fast directed ejections (e.g., jets) of plasma,
powerful flows of heat, and accelerated particles. They interact
with the chromosphere and photosphere, and therefore, creating an extremely
rich scenario of secondary physical processes observed as a
solar flares.

It is generally believed and well supported by observations
that magnetic reconnection is the key effect
which plays the crucial role in annihilating the complex magnetic field
structures and corresponding energy release.
The solar flares are mainly distinguished in two categories, e.g., the confined and
eruptive flares, which are usually triggered respectively in the closed and open morphology
of overlying magnetic fields.
The instabilities generated in the complex magnetic fields may be one of the most probable causes
to drive/trigger the solar flares after the reconnection of unstable flux tubes with the 
neighbourhood field configuration.
The emergence of unstable and helical twisted structures can trigger the flares followed
by an eruption (\citealt{liu2008}, 2007;  and references cited there). However, the activation of twisted helical magnetic structures may also
play a crucial role in the flare energy build-up and their initiation with
failed eruption depending upon the surrounding magnetic field environment
(\citealt{kumar2010b}, \citealt{sri2010} and references cited there).

Solar coronal loops may be considered as the current
($ \, \stackrel{<}{_\sim} \, $10$^{12}$ amp.) carrying conductors.
Two current carrying conductors possess net attractive force if both have resultant currents in the same
direction or resultant magnetic fields in the opposite direction depending upon their orientation with each other.
Collisions between current carrying loops are considered as a cause of some
solar flares \citep{sakai1996}.
Based on the loop orientations and size of the interaction region, the
current carrying loop interactions are classified into three categories:
(a) 1-D coalescence (I-type),
(b) 2-D coalescence (Y-type), and
(c) 3-D coalescence (X-type).
The theoretical model of \citet{gold1960} firstly explains the flare
triggering caused by interacting current carrying loops.
However, it is not necessary that the field lines should be anti-parallel
for the interaction of two current carrying conductors.
There may be other mechanisms, e.g., footpoint shear motion and rotation,
which can also destabilize the loop-system to trigger the flare and eruption.
Stronger shear has more probability for the initiation of the solar flares
and related eruptions (e.g., \citealt{tan2009} and references cited there).
Yohkoh has also observed some of the flaring events which show three types of loop
interaction (I, Y and X-type).
In the above mentioned interactions, the 3-D X-type reconnection due to coalescence is the
most realistic scenario in the active regions. The necessary condition for 3-D X-type interaction is that the length of the
interaction region (L) should be comparable to the loop diameter (R)
\citep{sakai1989}.

\citet{hana1996} has found evidence of the emergence of a small loop near
one of the footpoints of a pre-existing large coronal loop using observations
of various instruments including Yohkoh. The interaction of this loop
with the larger loop causes flares, micro\-flares and jets.
\citet{liu1998} have also observed the flare triggering by the I-type interaction
of loop-systems.
\citet{fale1999} have shown the flare energy release caused by two
successive X-type interaction of an expanding loop with two high-lying and
nearly parallel loop-systems.
Furthermore, \citet{poh2003} has also studied the series of flares from
AR 8996 on 18-20 May, 2000 and provided the evidence of flare triggering
due to loop-loop interaction with the observation of moving magnetic
features around the sunspot region.
 Several authors have reported the loop-loop interaction as a cause of solar flares. However,
further multiwavelength studies are needed to understand the flare triggering mechanism
due to loop-loop interaction, and its responses in the various layers of the solar atmosphere. 
In spite of the loop-loop interaction, the flare triggering followed by solar eruptions (e.g., coronal mass
ejection) can also be caused by the interaction of filaments system due to sunspot
rotation (e.g., \citealt{kumar2010a} and references cited there).

%
%%% Attention of the Authors !!! %%%   %%%%%%%%%%%%%%%%%%%%%%%%%%%%%%%%%%%%%%%
%
We know that the interacting current loops are not located in the vacuum or isolating 
medium, but they are lying in the highly-conducting plasma penetrated by frozen-in magnetic fields in the solar corona.
%On the one hand, we should not forget that the interacting current loops
%are located not in vacuum or isolating medium but in the highly-conducting
%plasma penetrated by frozen-in magnetic fields in the solar corona.
From the beginning of the evolution of a current carrying loop-system, 
every change in the current carrying loop-system generates currents in 
the surrounding plasma and magnetic field.
%Starting from appearance of a current loop or loops, every change in
%current-loop system generates currents in a surrounding plasma and magnetic
%field.
Therefore, we have to take into account an interaction not only between the loops but
also with these new currents, in particular with screening current layers
between the loops.
Moreover, the frozen-in magnetic fields of an active region or an activity
complex are typically strong in the corona and have their specific topology
determined by the photospheric sources.
 \citet{hen1987} were the first to show that these effects are
essential and must be considered in terms of magnetic reconnection of
field-aligned electric currents (see Section~\ref{sub:topology}).
%
%%% Attention of the Authors !!! %%%   %%%%%%%%%%%%%%%%%%%%%%%%%%%%%%%%%%%%%%%
%
On the other hand, if there were no current loops related with a twist of
magnetic flux tubes at all, even in this case, three-dimensional reconnection
between interacting magnetic fluxes
%%% in a quadrupole-type model %%%
gives such distribution of reconnected magnetic fluxes in the corona that
two soft X-ray loops look like %that they 
interacting with each other \citep{gor1989,gor1990}.
That is the reason that %why
the observations demonstrated such structures are usually
considered as a direct evidence of the hypothesis of two interacting currents.

In this paper, we present a multiwavelength study of M7.9/1N solar flare on 27 April, 2006 in AR NOAA 10875, which shows rare observational evidence of the coalescence and the interaction of two current carrying loops.
We report a most likely multiwavelength signature of X-type interaction and coalescence instability in the 
active region which triggers the solar flare. In Section~\ref{sub:observations}, we present multiwavelength observations of the event.
We discuss our results and conclusions in the last section.
%
%%%%%% Section 2 %%%%%%   %%%%%%%%%%%%%%%%%%%%%%%%%%%%%%%%%%%%%%%%%%%%%%%%%%%%
%
\section{OBSERVATIONS AND DATA}
   \label{sub:observations}

\begin{figure}
\centerline{
\includegraphics[width=10cm]{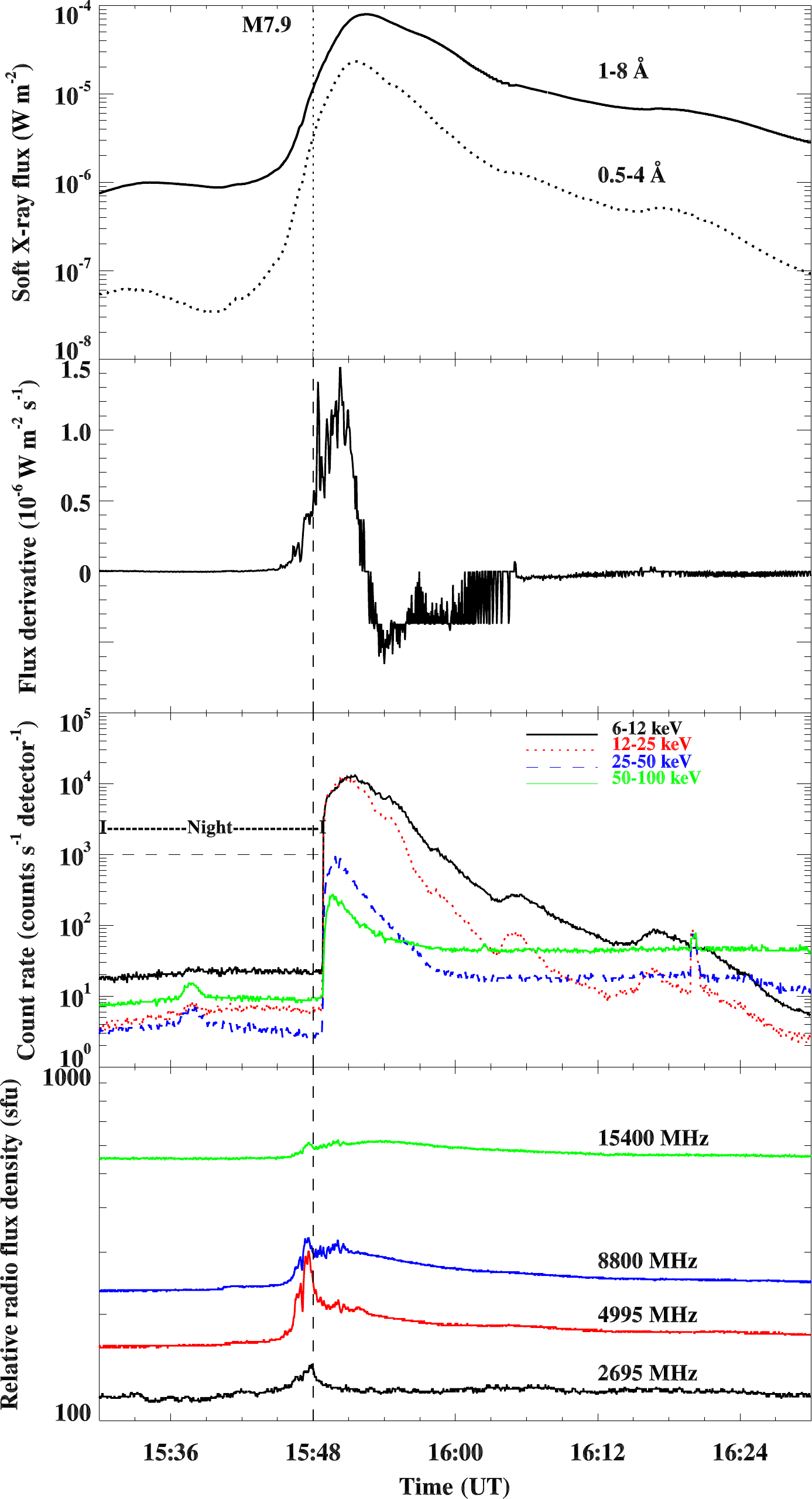}
}
\caption{Soft X-ray flux, flux derivative, RHESSI and radio flux  profiles
         for the M7.9 flare event on 27 April, 2006.
         The soft X-ray flux derivative matches well with the hard X-ray
         flux profile. This implies that the accelerated electrons that
         produce the hard-X-ray also heat the plasma that produces the
         soft X-ray (Neupert effect). The dotted line in the third panel indicates the RHESSI night time.}
   \label{fluxes}
\end{figure}

The active region NOAA 10875 was located at S10\,E20 on 27 April, 2006,
showing $\beta$$\gamma$/$\beta$$\gamma$$\delta$ magnetic configuration, and has produced M7.9/1N class solar flare.
According to the GOES soft X-ray flux profile, the flare started at 15:45~UT, was at its 
maximum at 15:52~UT and ended at 15:58~UT.
Figure \ref{fluxes}
%
%%% Fig. 1 %%%
%
displays the flux profiles in the soft X-ray, soft X-ray derivative, hard X-ray
and radio wavelengths. The flux derivative of soft X-ray matches well with the rise-up of hard
X-ray flux profile. This implies that the accelerated electrons that produce the hard X-ray
also heat the plasma that produces the soft X-ray, obeying the Neupert
effect \citep{neupert1968}.
%
%%% Attention of the Authors !!! %%%   %%%%%%%%%%%%%%%%%%%%%%%%%%%%%%%%%%%%%%%
%
More exactly, this means that the impulsive heating of the solar atmosphere
by accelerated electrons can dominate its heating by thermal fluxes from the
high-temperature source of flare energy (see Chapter 2 in Somov, 1992).
So there is a causal connection between the thermal and nonthermal flare
emissions. Further, the radio flux profile shows the sharp rise-up with double peak
structure mostly in 4.9 and 8.8 GHz at 15:47 UT, which shows the
gyro\-synchrotron emission generated by the accelerated electrons at the
reconnection (i.e. loop-interaction) site.

%
%%%%%% Sub-section 2.2 %%%%%%   %%%%%%%%%%%%%%%%%%%%%%%%%%%%%%%%%%%%%%%%%%%%%%
%
\subsection{GOES SXI AND TRACE Observations}
   \label{sub:GOES}

We have used GOES-SXI observations of the event \citep{hill2005,pizzo2005}.
It is a broadband imager in the 6--60~\AA \ bandpass that produces full-disk
solar images with $\sim$1 minute cadence.
The images consist of 512 pixel$\times$512 pixel with 5$^{\prime\prime}$ \
resolution.
The FWHM of the telescope point-spread function is $\sim$10$^{\prime\prime}$. A set of selectable thin-film entrance filters allows plasma 
temperature discrimination, i.e.,  open, three polyimide (thin, medium, and thick), and three beryllium (thin, medium, and thick). The open and polyimide filters are sensitive 
to the plasma below 2 MK.  It is especially suitable for continuous tracking of coronal loops.

\begin{figure}
\centering{
\hspace{3.0 mm}
\includegraphics[width=5.9cm]{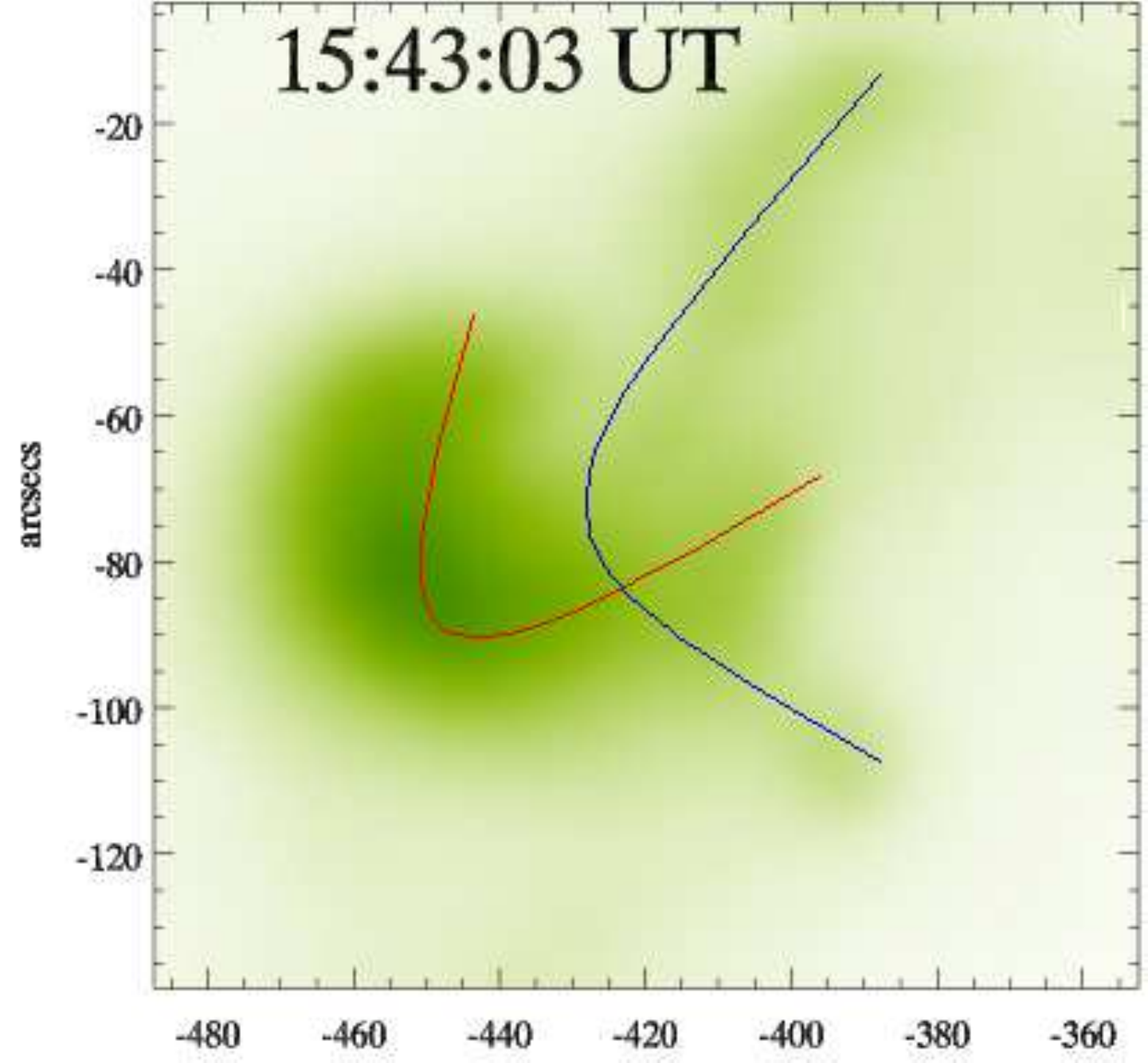}
\includegraphics[width=5.5cm]{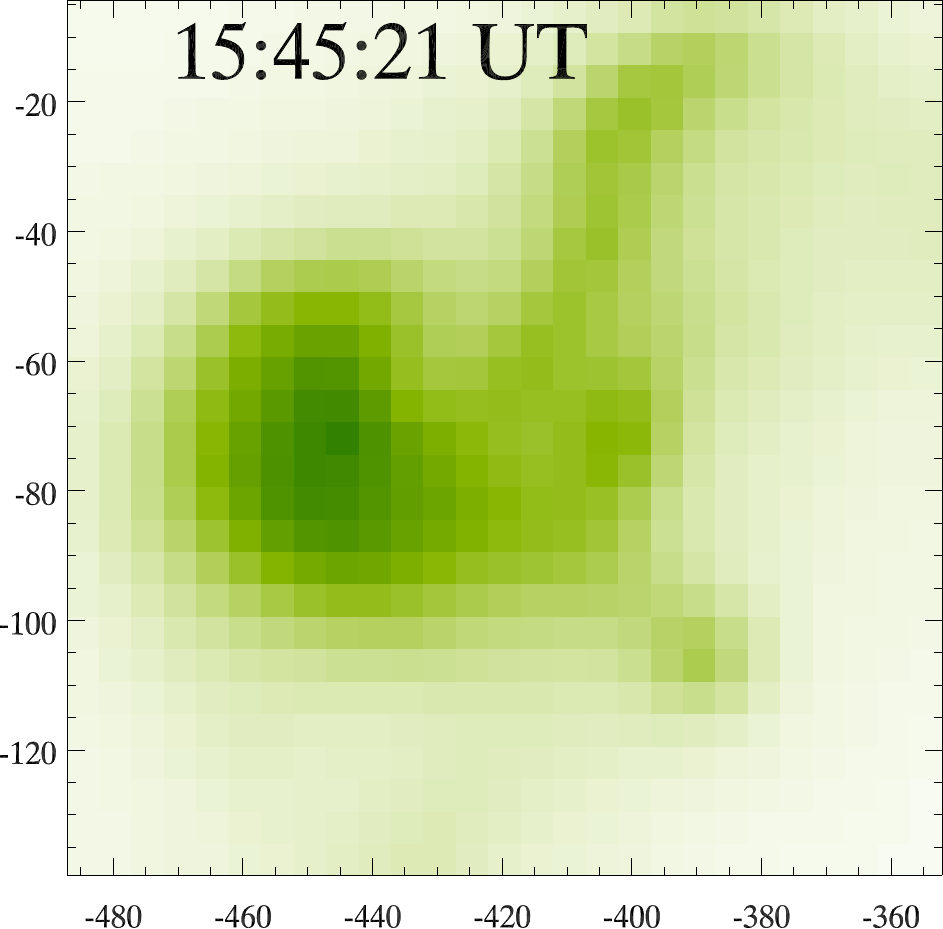}

\hspace{3.0 mm}
\includegraphics[width=5.8cm]{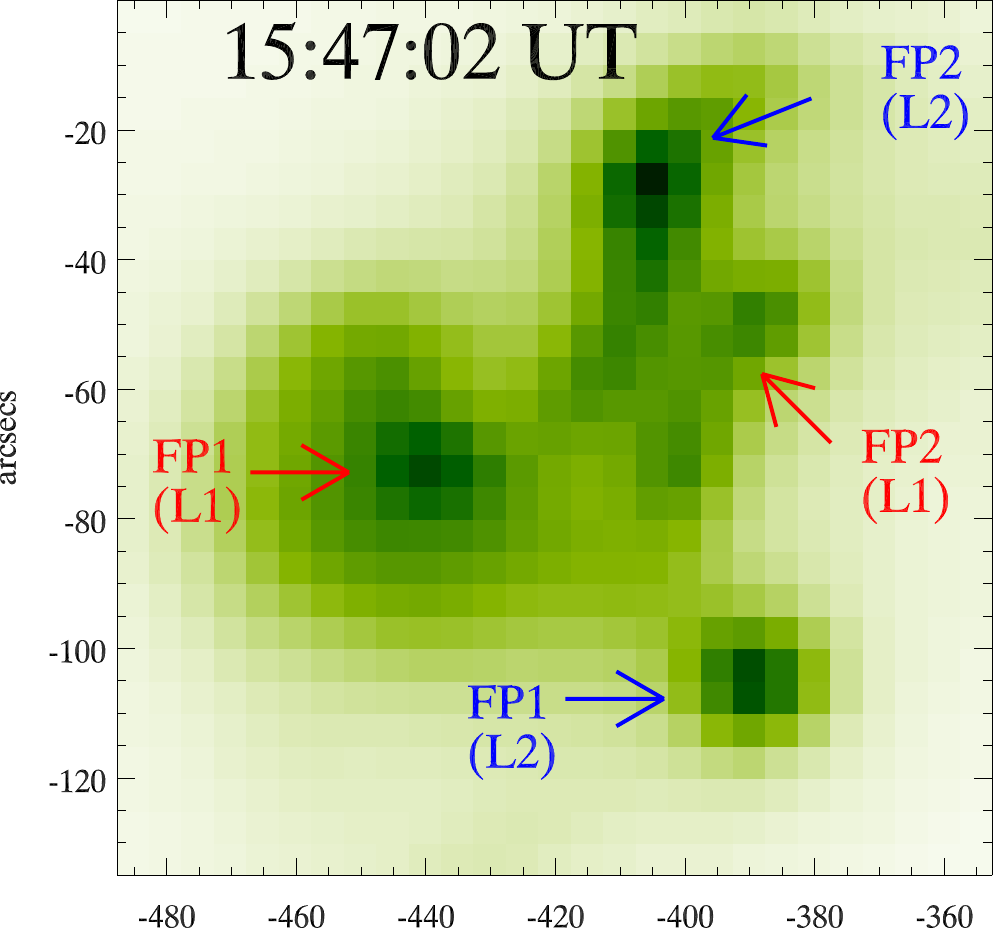}
\includegraphics[width=5.6cm]{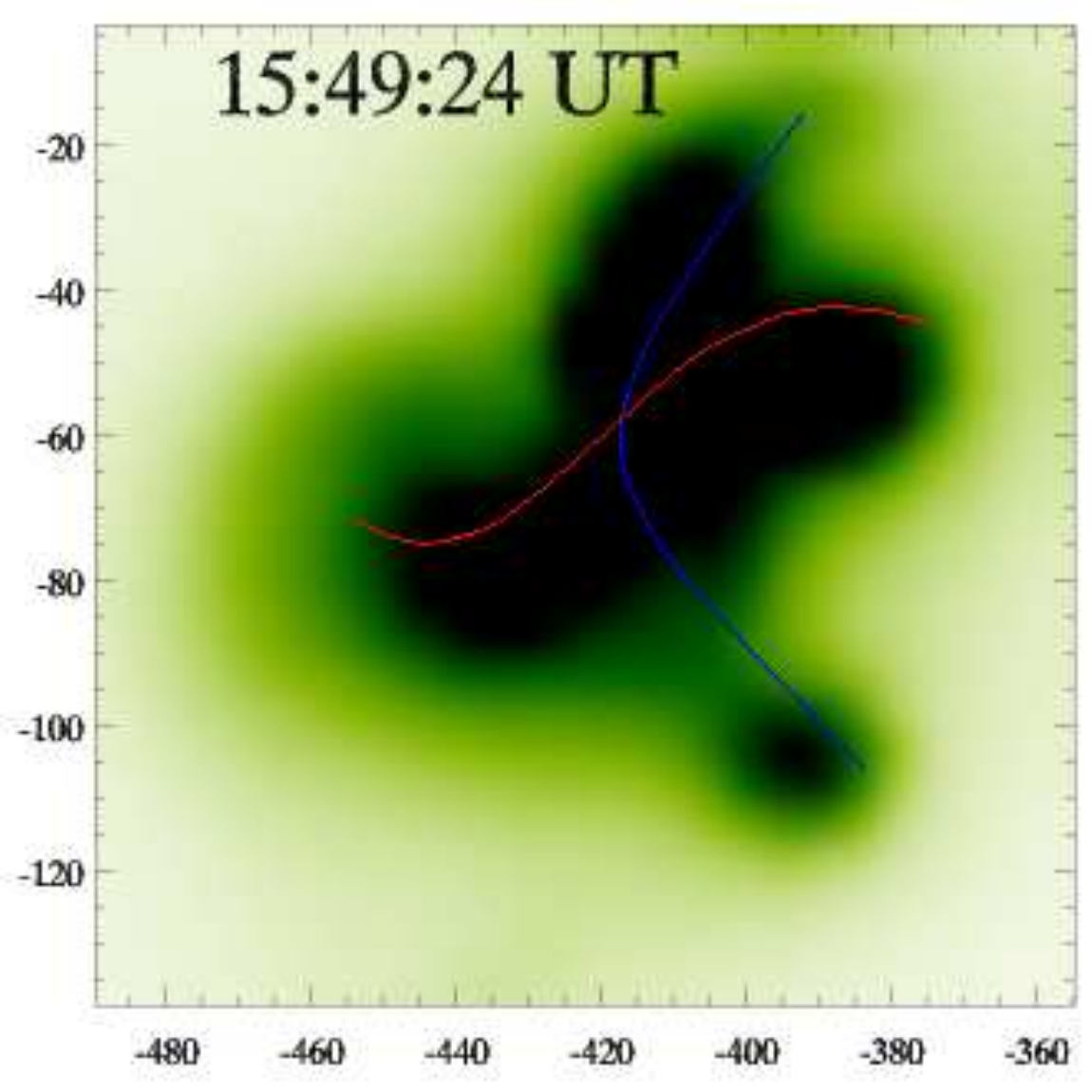}

\hspace{3.0 mm}
\includegraphics[width=5.8cm]{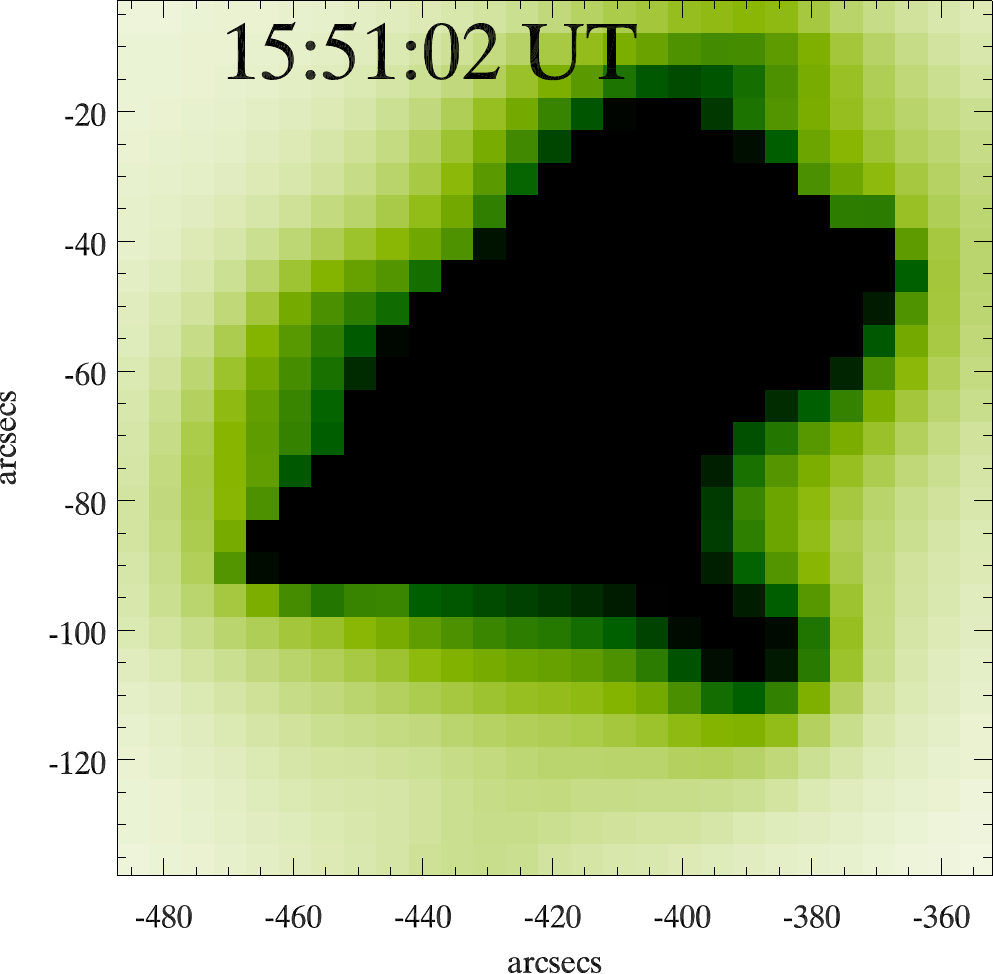}
\includegraphics[width=5.6cm]{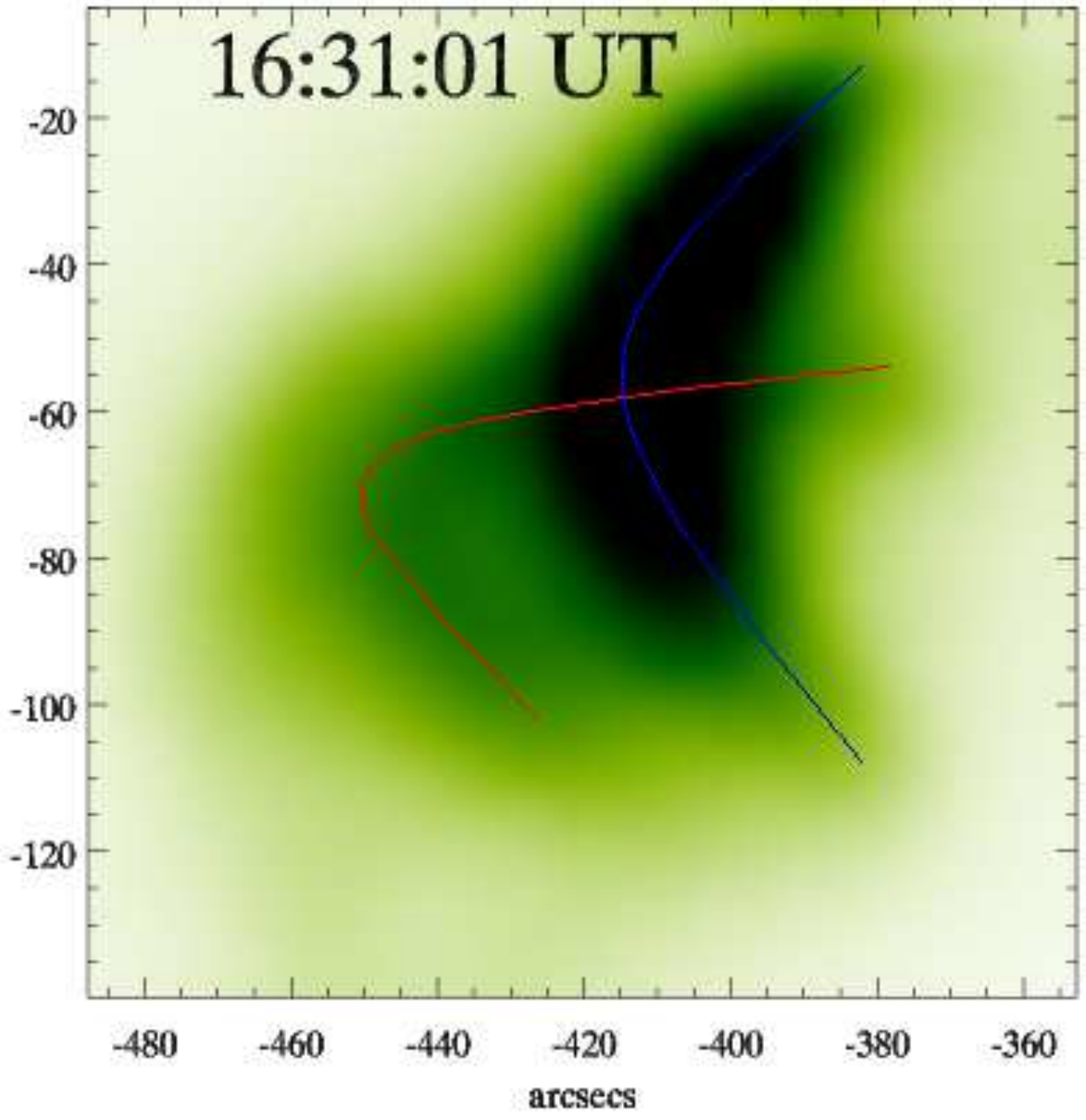}
}
\caption{GOES Soft X-ray coronal negative images (6--60~\AA)
         showing the flare evolution with the interaction of two coronal
         loops on 27 April, 2006. The upper left panel shows a
         lower loop system (blue) underlying a higher loop system (red).
         The lower loop first looks brighter during flare initiation.
         The middle left panel shows the corresponding footpoints of both interacting
         loops indicated by FP1 (L1) and FP2 (L1) for loop 1 and FP1 (L2) and 
	 FP2 (L2) for loop 2, respectively. 
         The bottom left panels shows the flare maximum due to loop-loop
         interaction and the bottom right panel indicates the simplified 2
         loops after the flare energy release.}
   \label{sxi}
\end{figure}

Figure \ref{sxi}
%
%%% Fig. 2 %%%
%
displays the selected images of GOES SXI before and during the flare 
activity. 
Two loop systems have been observed before the flare initiation. One lower loop system (indicated by red line)  
is underlying a higher loop-system (blue). 
Initially, brightening starts in the lower loop during flare initiation 
at 15:43~UT.  
%
%%% Attention of the Authors !!! %%%   %%%%%%%%%%%%%%%%%%%%%%%%%%%%%%%%%%%%%%%
%
This loop becomes more brighter as the flare progresses.  
The four foot\-points of  both the loop-systems become evident at 
15:47~UT 
%
%%% Attention of the Authors !!! %%%   %%%%%%%%%%%%%%%%%%%%%%%%%%%%%%%%%%%%%%%
%
mainly due to the  precipitation of the accelerated electrons from the interaction 
or reconnection site. The corresponding footpoints of both interacting
loops are indicated by FP1 (L1) and FP2 (L1) for loop 1 and FP1 (L2) and 
FP2 (L2) for loop 2, respectively. 
As the plasma is heated-up due to the dissipation of kinetic energy of the accelerated electrons from the reconnection site,
%
%%% Attention of the Authors !!! %%%   %%%%%%%%%%%%%%%%%%%%%%%%%%%%%%%%%%%%%%%
%
chromospheric evaporation takes place and it fills the interacting loop-system 
%
%%% Attention of the Authors !!! %%%   %%%%%%%%%%%%%%%%%%%%%%%%%%%%%%%%%%%%%%%
%
in the corona and these loops look like as if they are crossing to each other. 
Now the X-type configuration becomes evident at 15:49 UT.  
The flare maximum takes place at 15:52~UT. 
After the interaction between the loops, the orientation of the lower loop 
has changed into a more relaxed state. 
The SXI image taken during the decay phase of the flare (at 16:31~UT) 
evidently shows the orientation change of the lower loop-system. 

In this Figure, the loop shown by red line is marked in the upper-left
panel as rooted somewhere close to X$\approx$-445$^{\prime\prime}$, Y$\approx$-50$^{\prime\prime}$.
However, in middle-left panel the left foot of this loop
(marked by FP1(L1) has co-ordinates at X$\approx$-440$^{\prime\prime}$, Y$\approx$-70$^{\prime\prime}$. Therefore, the shift in the footpoint during the dynamical flare event is $\Delta$X = 5$^{\prime\prime}$, $\Delta$Y = 20$^{\prime\prime}$. Presumably, this apparent
displacement of the footpoint FP1(L1) may be due to the two reasons:

(a) A displacement directed out from the photospheric neutral
    line, therefore, it is related to the motion of the flare ribbons
    in the opposite directions. Such behavior is typical for the
    two-ribbon flares;

(b) A displacement directed parallel to the photospheric neutral
    line, which is related to the magnetic shear relaxation.

These two processes can jointly cause an increasing or decreasing
distance between the footpoints. Investigations in the frame of a more detailed model
should be done to interpret this feature. It is necessary
to compare the kernel displacements observed during the flare
with motions and evolution of magnetic fields in the photosphere
before the flares (see Somov et al., 2002).

\begin{figure}
\centering{
\hspace{3.0 mm}
\includegraphics[width=4.88cm]{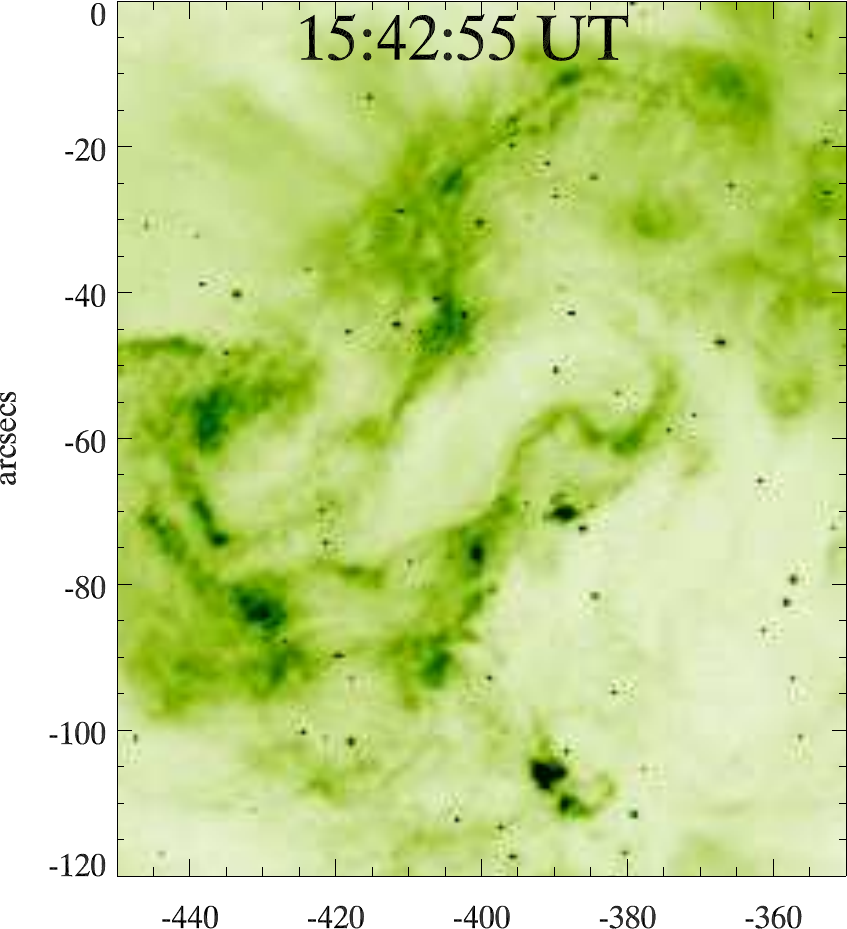}
\includegraphics[width=4.6cm]{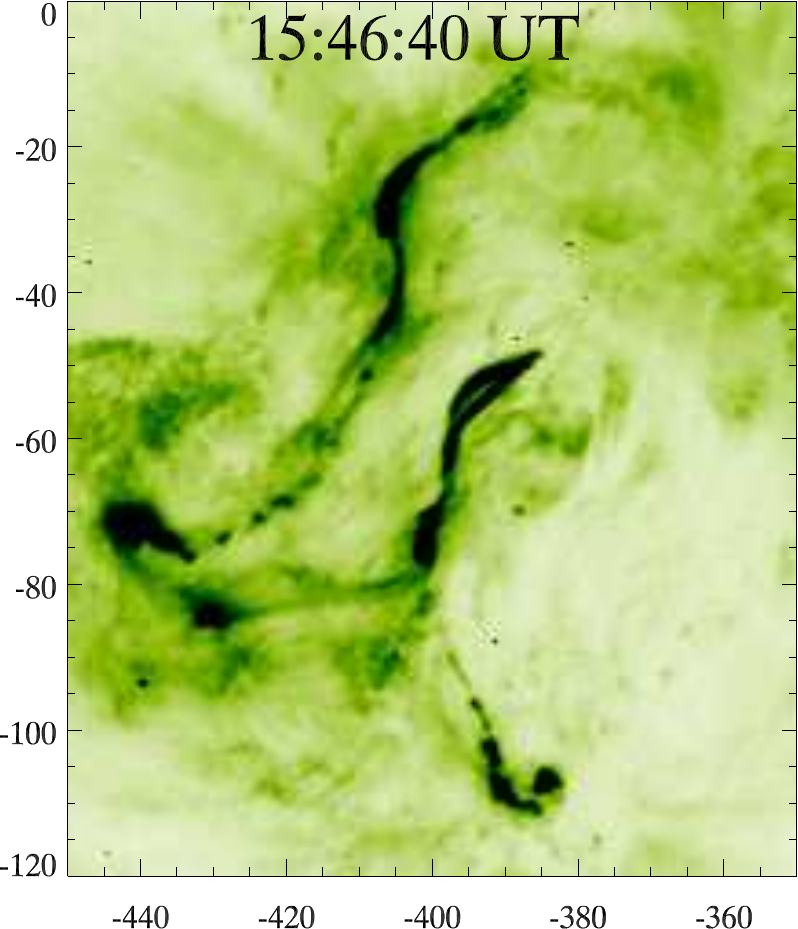}
\includegraphics[width=4.6cm]{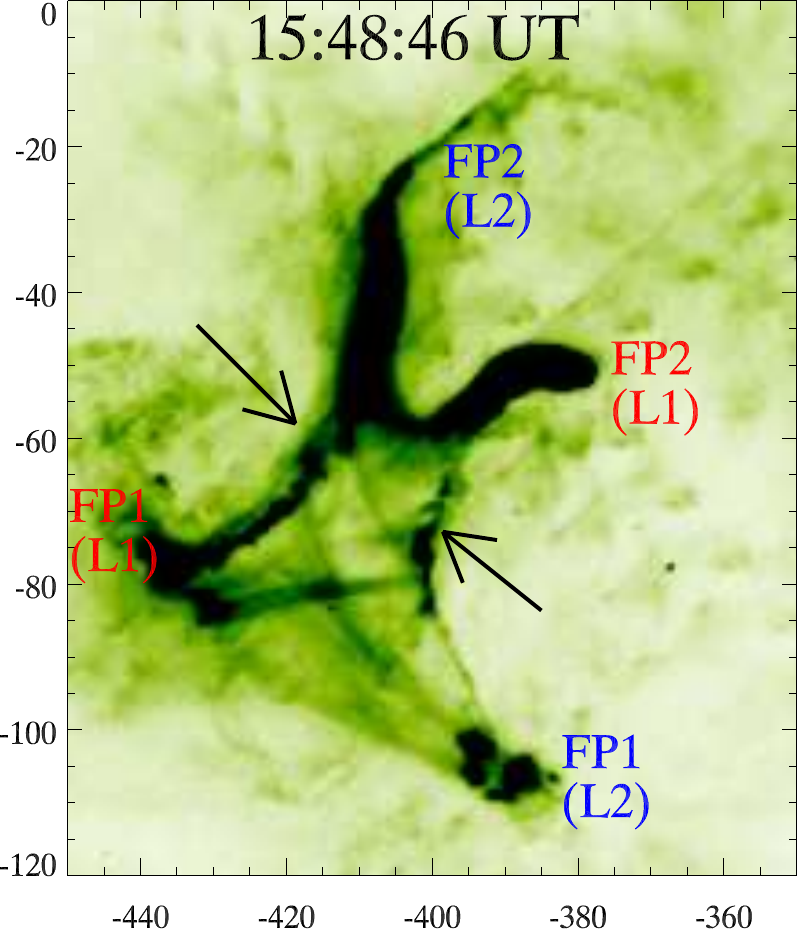}

\hspace{3.0 mm}
\vspace{1.0 mm}
\includegraphics[width=4.88cm]{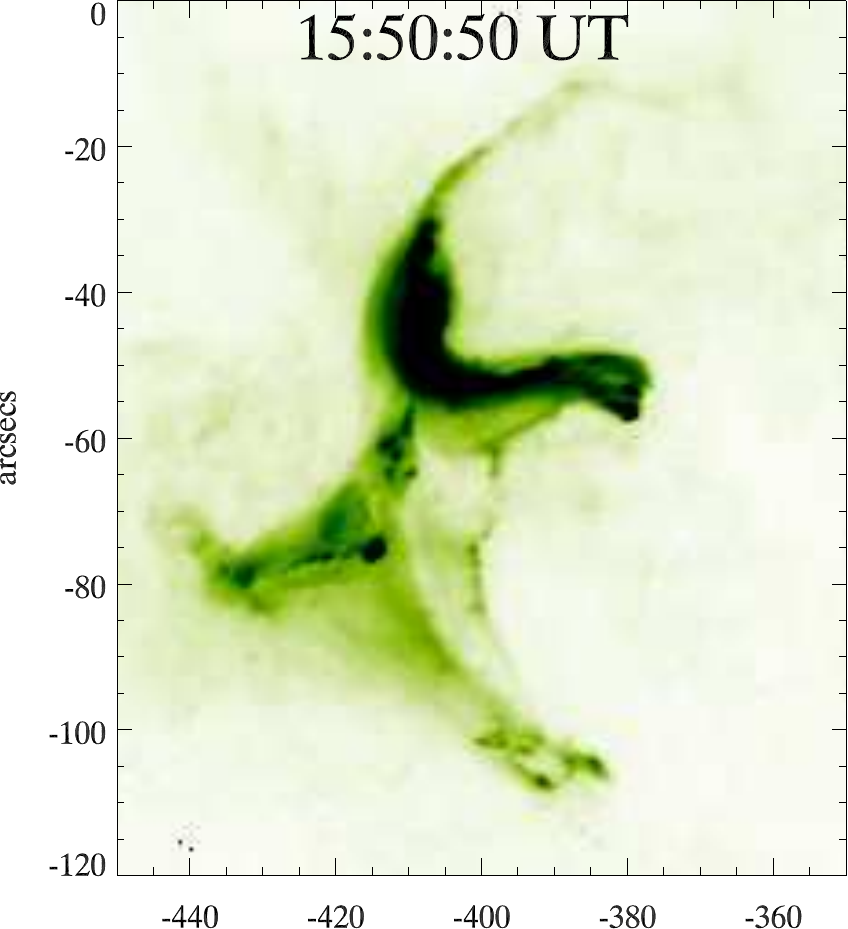}
\vspace{1.0 mm}
\includegraphics[width=4.6cm]{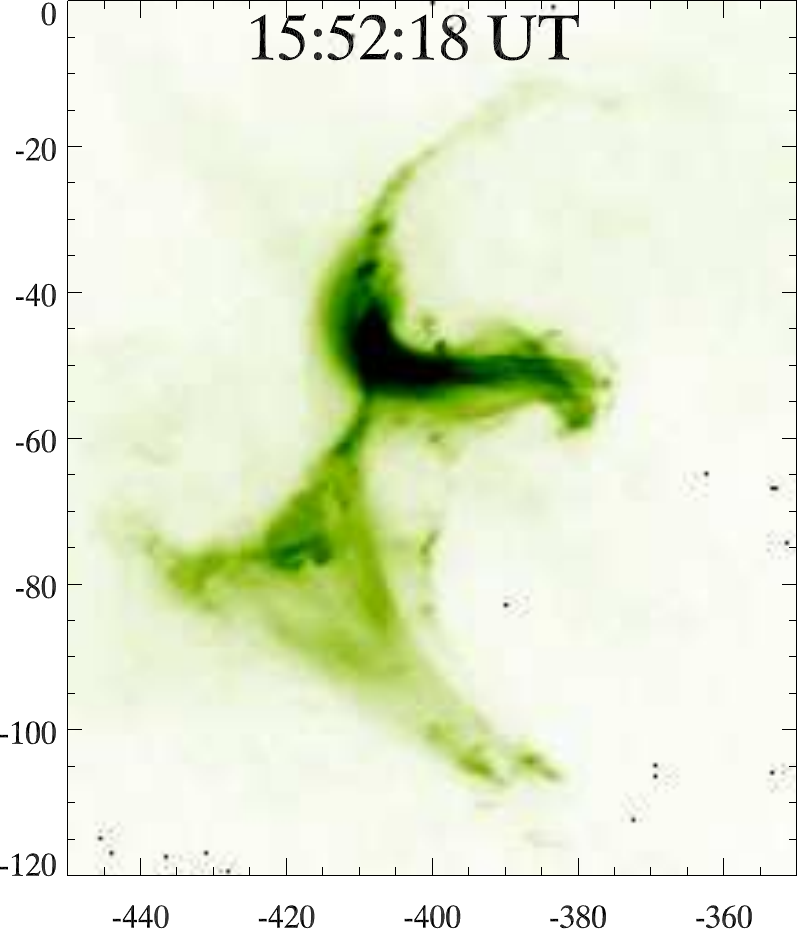}
\vspace{1.0 mm}
\includegraphics[width=4.6cm]{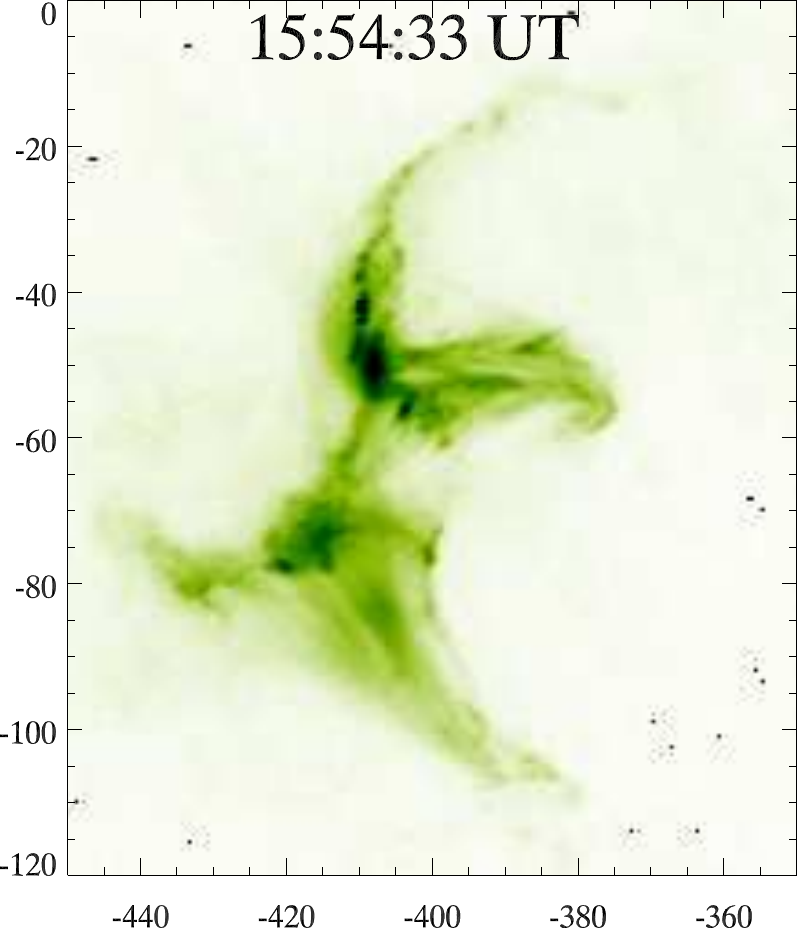}

\hspace{3.0 mm}
\vspace{1.0 mm}
\includegraphics[width=4.88cm]{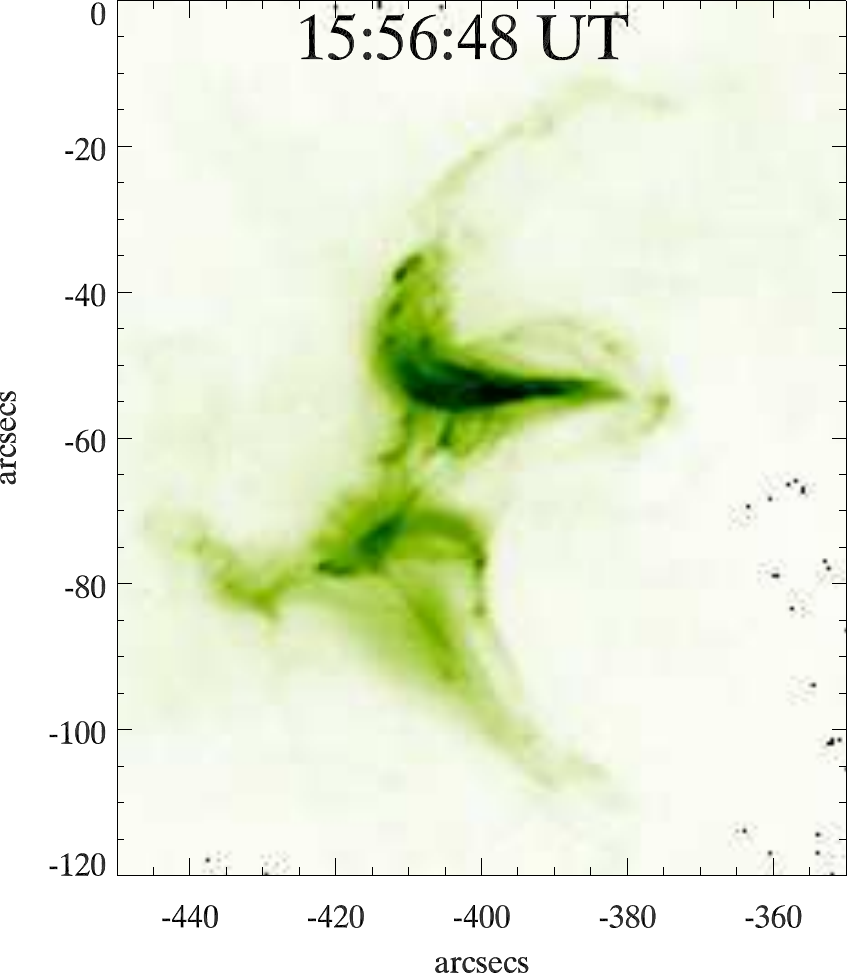}
\includegraphics[width=4.6cm]{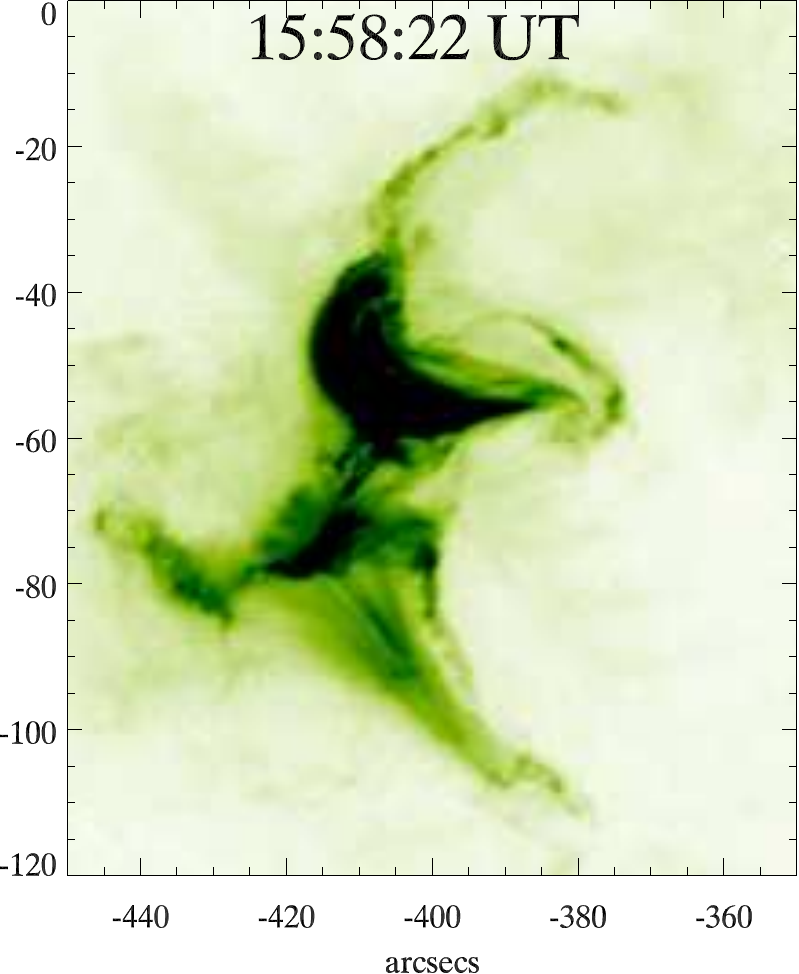}
\includegraphics[width=4.6cm]{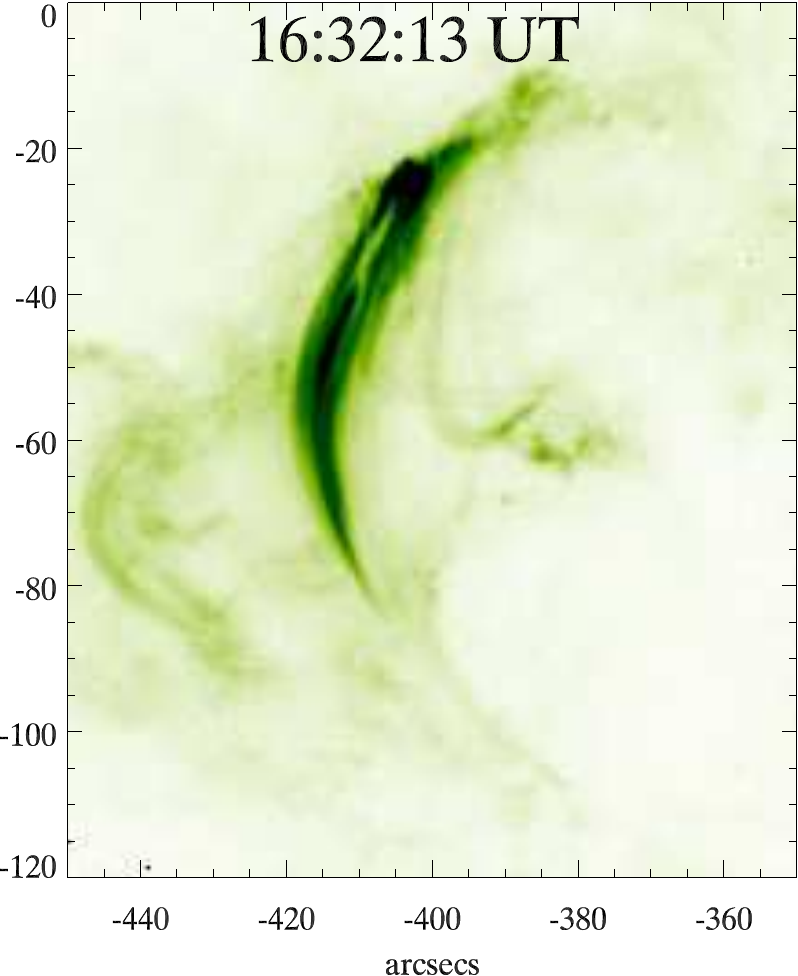}
}
\caption{TRACE 195 \AA \ negative images showing the flare
         evolution with the interaction of two coronal loops on 27 April,
         2006. The upper  and middle panels show approaching and
         interacting loops.
         The flare initiation takes place as the loops approach and
         maximizes at the time of interaction. The corresponding footpoints of the interacting loops are indicated by FP1 (L1) and FP2 (L1) for loop 1 and FP1 (L2) and FP2 (L2) for loop 2 respectively. The arrows indicate the interaction region/reconnection site. The bottom right panel shows the
         relaxation and orientation changes of the loops after interaction.
        }
   \label{tr_195}
\clearpage
\end{figure}

\begin{figure}
\centerline{
\includegraphics[width=12cm]{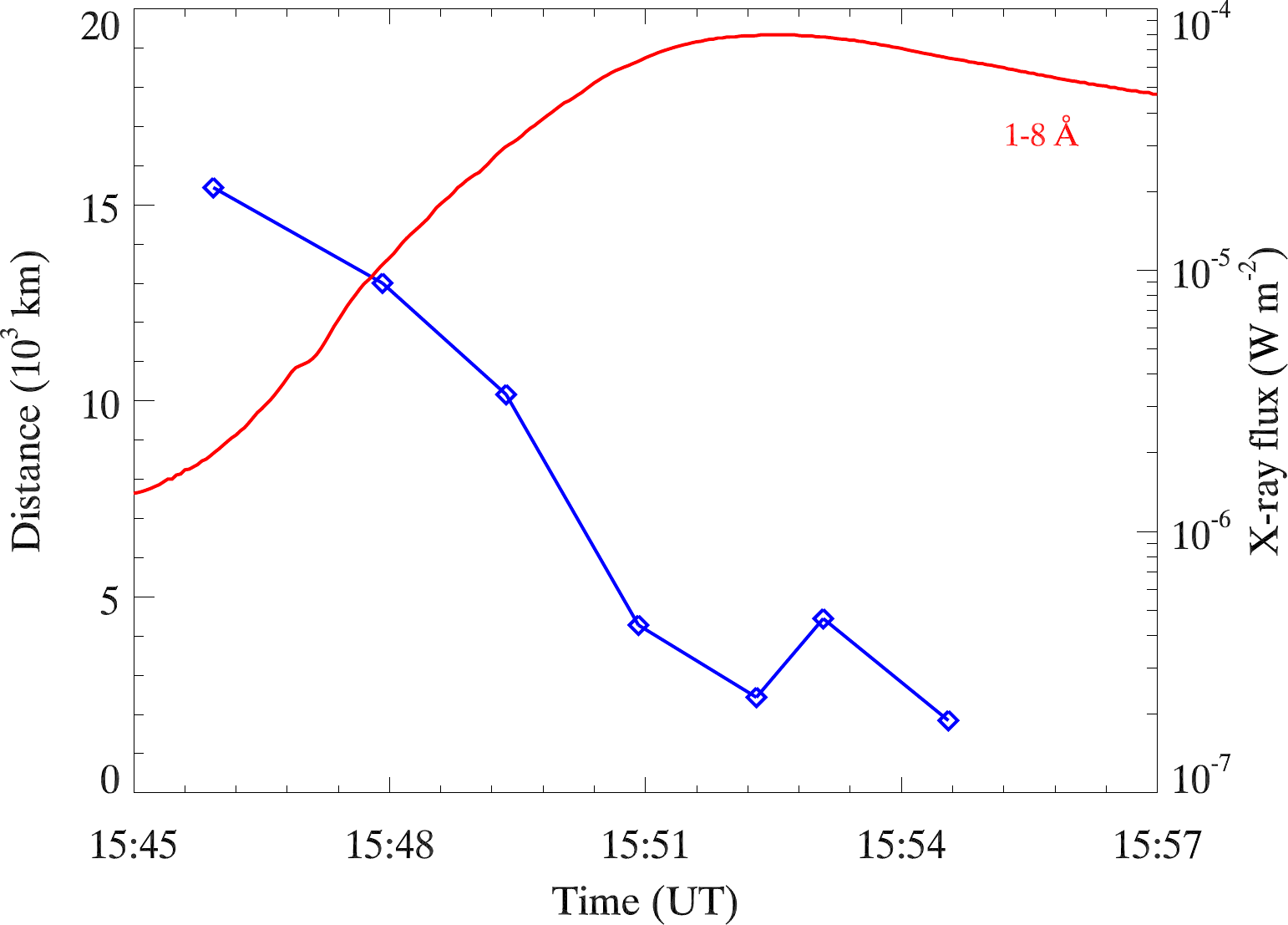}
}
\caption{The thickness of interaction region shown by blue curve (estimated from TRACE 195 \AA \
         images) plotted against GOES soft X-ray flux profile (red curve).
         This plot reveals that as the thickness of interaction region
         decreases, the soft X-ray flux increases.
         This may be the most likely signature of ongoing reconnection at the site of
         loops-interaction.
         The typical converging speed of interacting region is $\sim$30 km s$^{-1}$.}
   \label{thick_xray}
\end{figure}

TRACE (Transition Region and Coronal Explorer) provides the opportunity
to observe the Sun from chromosphere to corona \citep{handy1999}.
We have used TRACE 195~\AA \ (Fe XII, T$\sim$1.5 MK) and
1600~\AA \ (T$\sim$4000-10000 K).
The field of view for each image is 1024$\times$1024 with 0.5$^{\prime\prime}$ \ pixel$^{-1}$ resolution. The  typical cadence for TRACE images is $\sim$20-60 sec.
Figure \ref{tr_195}
%
%%% Fig. 3 %%%
%
displays the selected TRACE 195~\AA \ images during the flare activity.
TRACE data have been calibrated and analyzed using standard routines in the 
solarsoft library \footnote [2] {http://hesperia.gsfc.nasa.gov/ssw/trace/}.
During the flare initiation, brightening was observed along both sides
of the photospheric neutral line.
Two bright sheared structures are observed at 15:46 UT.
The image at 15:48 UT shows the loop-loop interaction and formation of an `X' point in
between the interacting loop-system.
Many interacting small flux threads/tubes may be seen in this image.
 After the X-type interaction during the impulsive phase of the flare, it seems that
the loop threads are changing their footpoint connectivities. This is the signature of
 an ongoing reconnection process in the same global configuration of the active region.
During 15:42--15:46 UT, the two interacting loops
are visible in the soft X-ray GOES/SXI images, however, they are not visible
in the TRACE images of the same duration. The GOES/SXI images represent the high temperature
and high coronal part of the loop systems, while the TRACE images show the lower part of the 
loop systems joining the two brightened ribbons. In the pre-flare state, the GOES/SXI images
show the loop segments visible due to the emission of the soft X-ray during loop-loop interaction, while at the same time
the plasma at EUV temperature band is not uploaded in the lower segments of the two loops to brought them as visible
as GOES/SXI images. However, near the flare maximum and even after the flare, the interacting loop systems are
clearly evident in both X-ray as well as in EUV, and imply the presence of plasma at various temperatures.
Since,  we see the different segments of the interacting loop-systems in GOES/SXI and TRACE images. Therefore, they
look like with a different orientations as the apex part may be more tilted compared to the lower segments.
%
%%% Attention of the Authors !!! %%%   %%%%%%%%%%%%%%%%%%%%%%%%%%%%%%%%%%%%%%%
%
We can identify the four foot\-points of the associated interacting
two loop-systems.
%
%%% Attention of the Authors !!! %%%   %%%%%%%%%%%%%%%%%%%%%%%%%%%%%%%%%%%%%%%
During the interaction time, the thickness of the interaction region (indicated by arrows)
reduces during the impulsive phase of the flare and it seems that the
orientation of the loops is changed during the flare maximum
(refer to image at 15:50~UT and onwards images).
During the sharp impulsive phase, the foot\-points of the loop systems do not show
significant changes (see TRACE movie). It means that the reconnection point is mostly fixed,
i.e., the loops interaction site.
The loop-system morphology becomes simple and relaxed during the decay phase of
the flare as observed in SXI images (see SXI image at 16:31:01 UT).
%
%%% Missing definition %%%
%%% Attention of the Authors !!! %%%   %%%%%%%%%%%%%%%%%%%%%%%%%%%%%%%%%%%%%%%
%
The thickness of the interaction region
%
%%% Show it by arrows in Fig. 4 %%%
%
is plotted against the GOES soft X-ray flux profile (refer to
Figure \ref{thick_xray}).
%
%%% Fig. 4 %%%
%
This plot reveals that the X-ray flux rises up as the thickness of the interaction region
decreases. This may be the most likely signature of ongoing reconnection at the loops interaction site.
From the linear fit, the typical converging speed is estimated as $\sim$30 km s$^{-1}$.
This speed may be related with the typical inflow speed as observed in other flares \citep{tsun1997,yok2001}.

\begin{figure}
\centering{
\includegraphics[width=7.5cm]{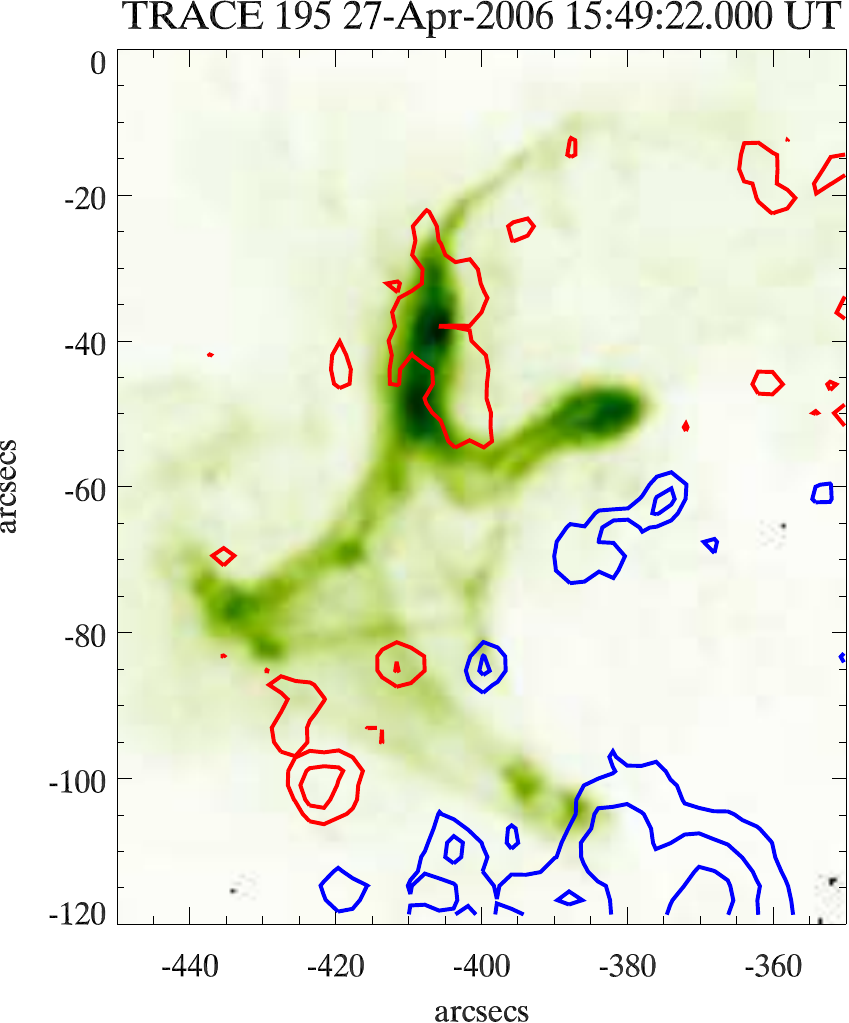}
\includegraphics[width=7.5cm]{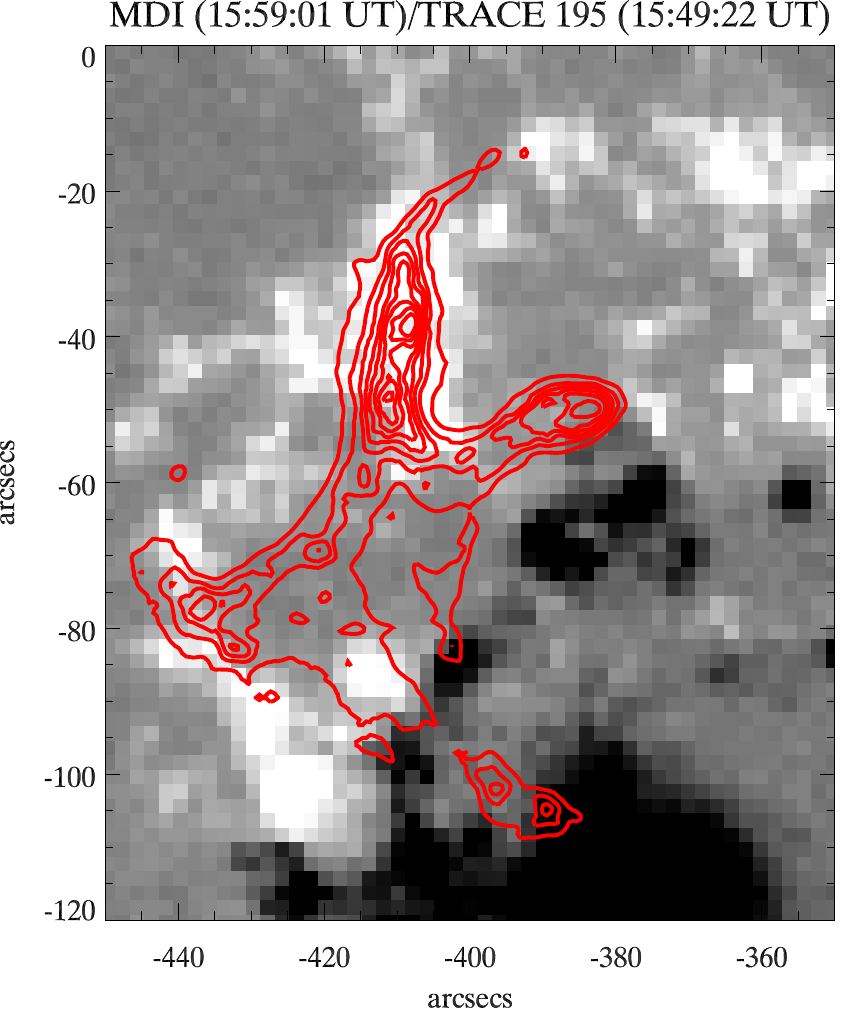}
}
\caption{Left: MDI contours overlaid on TRACE 195~\AA \ images during flare
         maximum (Blue contours indicate the negative whereas red contours
         show the positive polarity sunspots). The contour levels are $\pm 500, 
\pm 1000, \pm 2000, \pm 3000$ G.
         Right: TRACE 195 \AA \ contours overlaid on MDI magnetogram
         (Black=negative, White=positive).}
   \label{tr_mdi}
\end{figure}

We have overplotted MDI contours over TRACE 195 \AA \ image
and vice versa (refer to Figure \ref{tr_mdi}).
%
%%% Fig. 5 %%%
%
Left foot\-points [FP1(L1) and FP2(L2)] of the associated loop-systems are anchored in positive
polarity field regions whereas the right foot\-points [FP1(L2) and FP2(L1)] are anchored in the
negative polarity regions.

\begin{figure}
\centering{
\includegraphics[width=7.5cm]{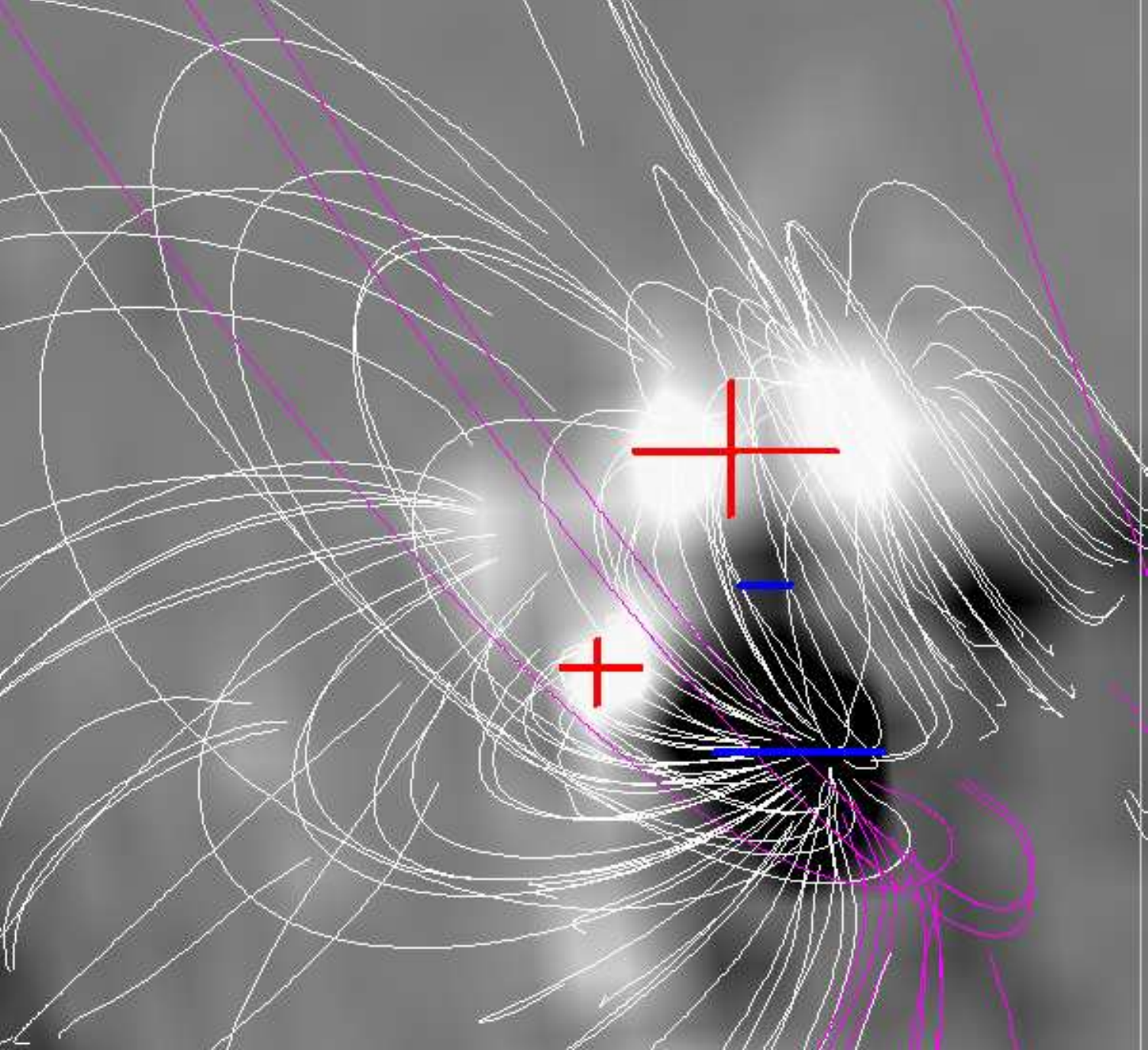}
\includegraphics[width=7.1cm]{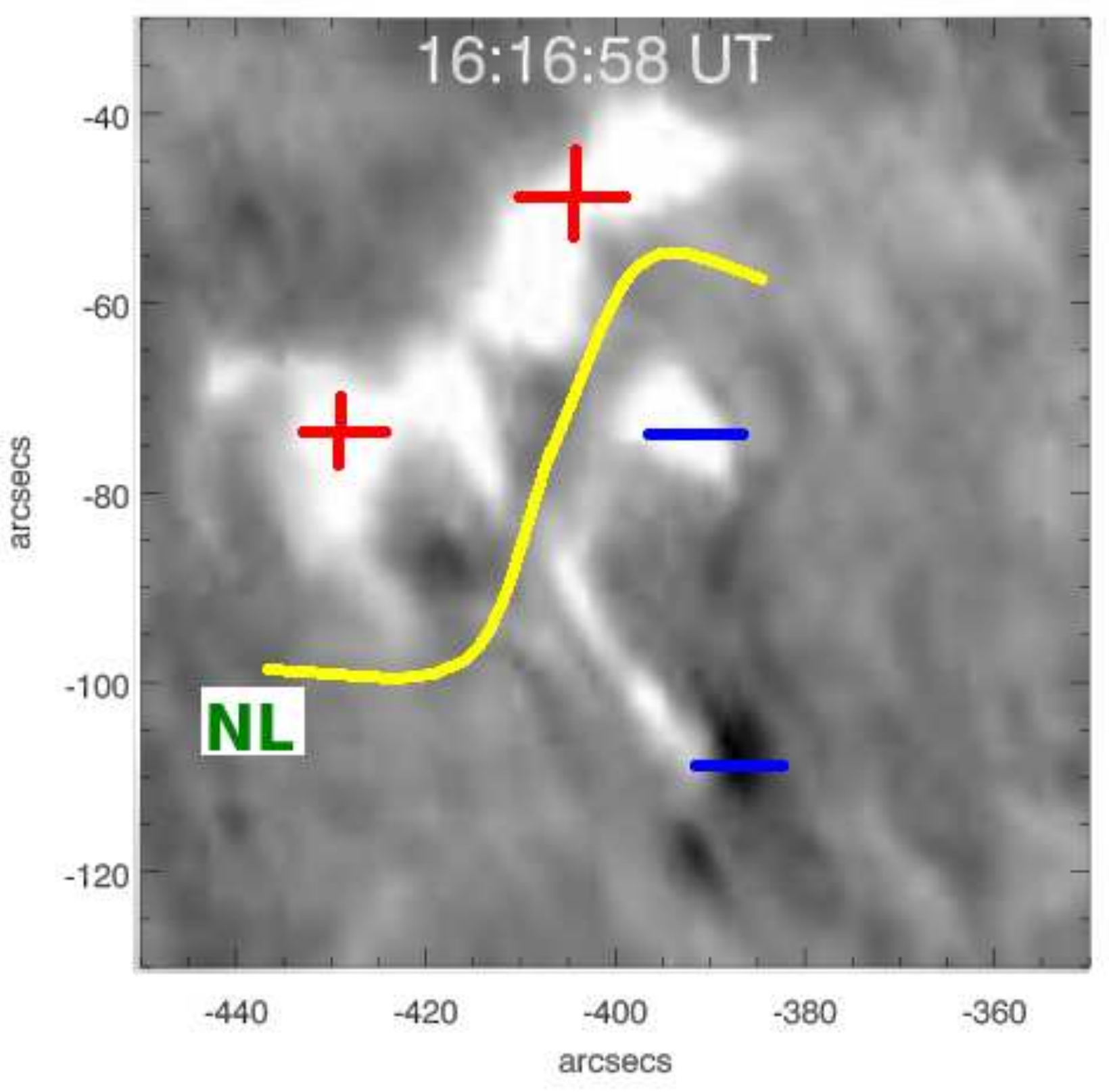}
}
\caption{Left: PFSS (Potential Field Source Surface) extrapolations using SOHO/MDI magnetogram at 00:05:00 UT on 27 April, 2006.
         Right: H$\alpha$ image during the decay phase of the flare showing
         flare ribbons on the both side of neutral line (NL), indicated by yellow line. The polarity at the location of flare ribbons is indicated by `+' and `--' symbols. For comparison, the locations of the flare ribbon polarities are denoted by  `+' (red) and `-' (blue) signs in the left panel.}
   \label{extr_ha}
\end{figure}

\begin{figure}
\centering
{
\includegraphics[width=7cm]{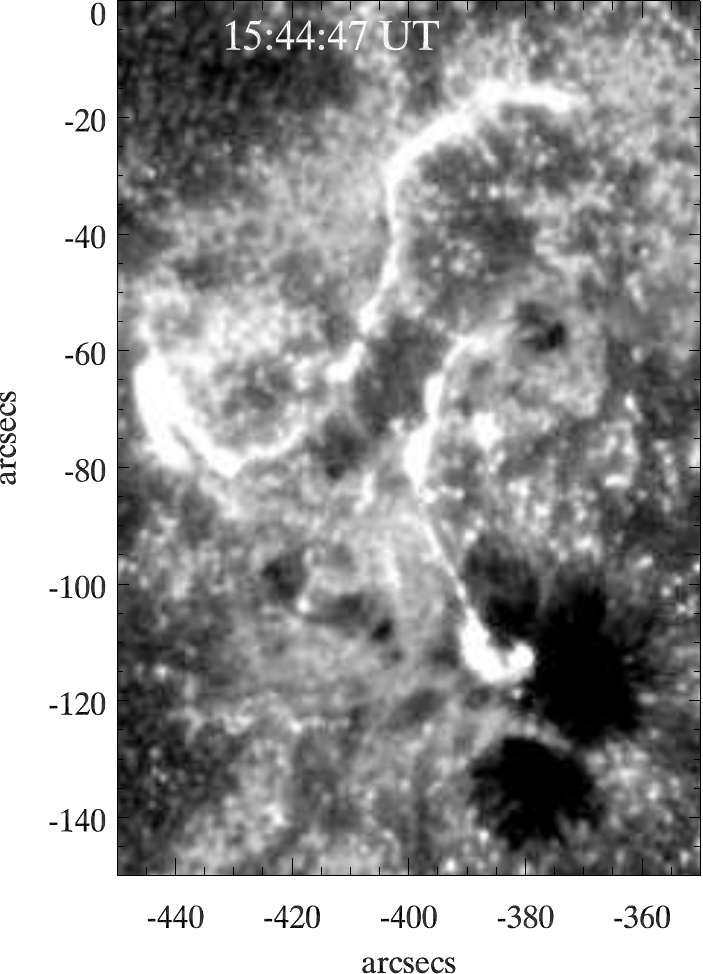}
\includegraphics[width=7.1cm]{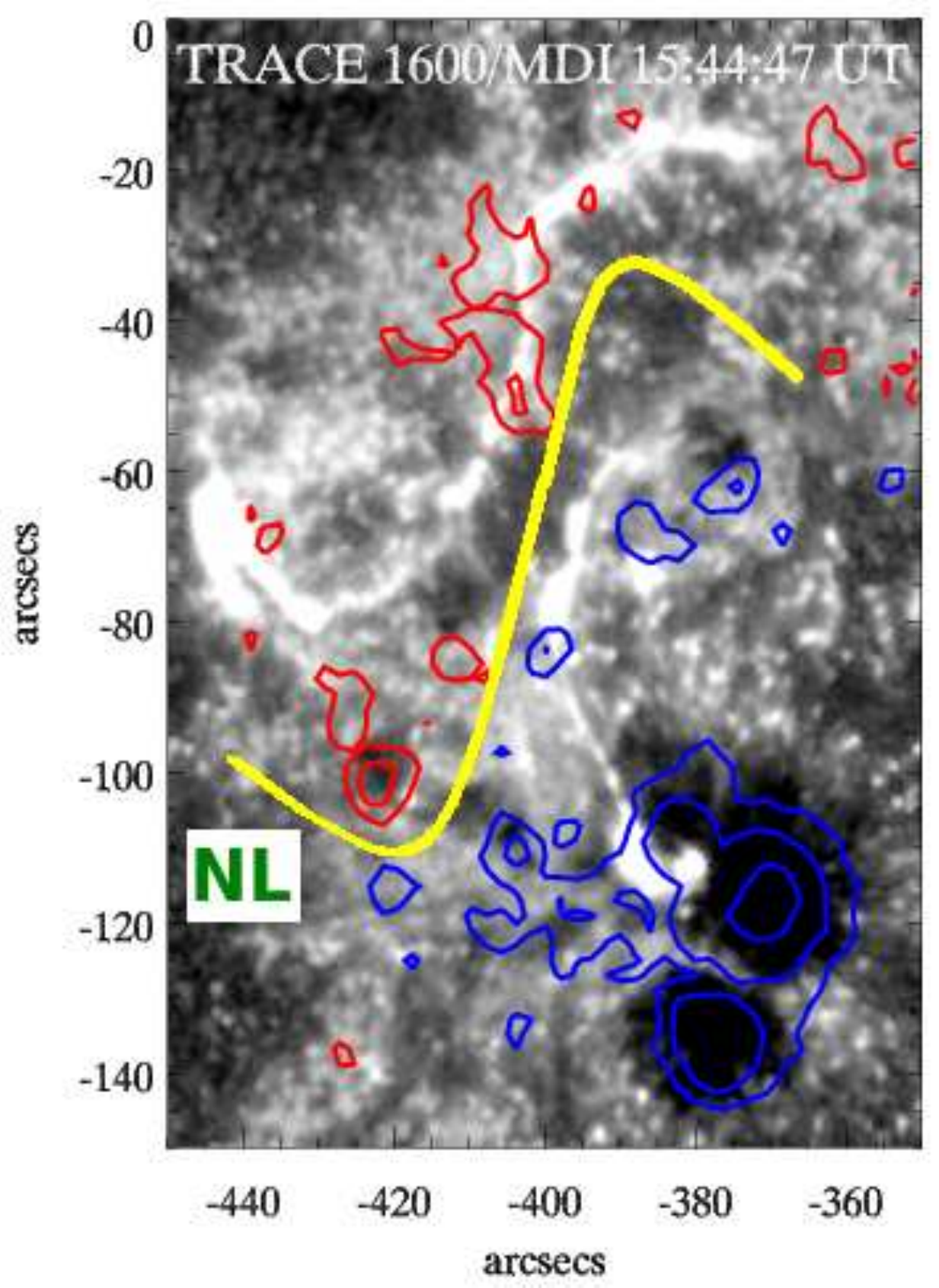}
}
\caption{ Left: TRACE 1600 \AA \ images showing the the morphology of flare
         ribbons during the flare. Right: SOHO/MDI magnetic field contours overlaid on TRACE 1600 \AA \ image. Red one indicate the positive polarity whereas blue one show the negative polarity fields. The contour levels are $\pm 500, \pm 1000, \pm 2000, \pm 3000$ G. Ribbons are formed on the both sides of neutral line (NL), drawn by yellow color.}
   \label{tr_rib}
\end{figure}

For investigating the overlying magnetic field environment of this active
region, we have used the potential field source surface (PFSS) extrapolation
\citep{alt1969,sch1969} before the flare event at 00:05 UT
(see left panel of Figure \ref{extr_ha}).
The right panel of Figure \ref{extr_ha} displays H$\alpha$ image observed at Meudon, which shows flare ribbons during the decay
phase (at 16:16~UT) of the flare.
%
%%% Fig. 6 %%%
%
It shows  mainly four bright kernels, which are the regions where most of
the energy flux is concentrated i.e. the sites of particle precipitation.
These are the footpoints of the corresponding reconnecting loop-system.
These observations are in favour of loop-loop interaction mechanism.  For comparison, the location of the flare ribbons polarities is 
denoted by corresponding `+' (red) and `-' (blue) signs in 
SOHO/MDI image of the active region (AR10875) with its coronal field extrapolation.
%
%%% Fig. 6 %%%
%
The coronal magnetic field topology is on average also in agreement with TRACE
and SXI observations.
Figure \ref{tr_rib}
%
%%% Fig. 7 %%%
%
displays the TRACE 1600~\AA \ images during the flare event.
Two ribbons, located on the both side of neutral line
are observed at 15:44 UT.
Left side ribbon shows the sheared `S' shaped structure, whereas the
ribbon at the right side shows simple structure.

%*****************************************************************************
%%% Sub-section 2.2 %%% %%%%%%%%%%%%%%%%%%%%%%%%%%%%%%%%%%%%%%%%%%%%%%%%%%%%%%
%*****************************************************************************
\subsection{Radio and RHESSI Observations}
   \label{sub:radio}

\begin{figure}
\centerline{
\includegraphics[width=9cm]{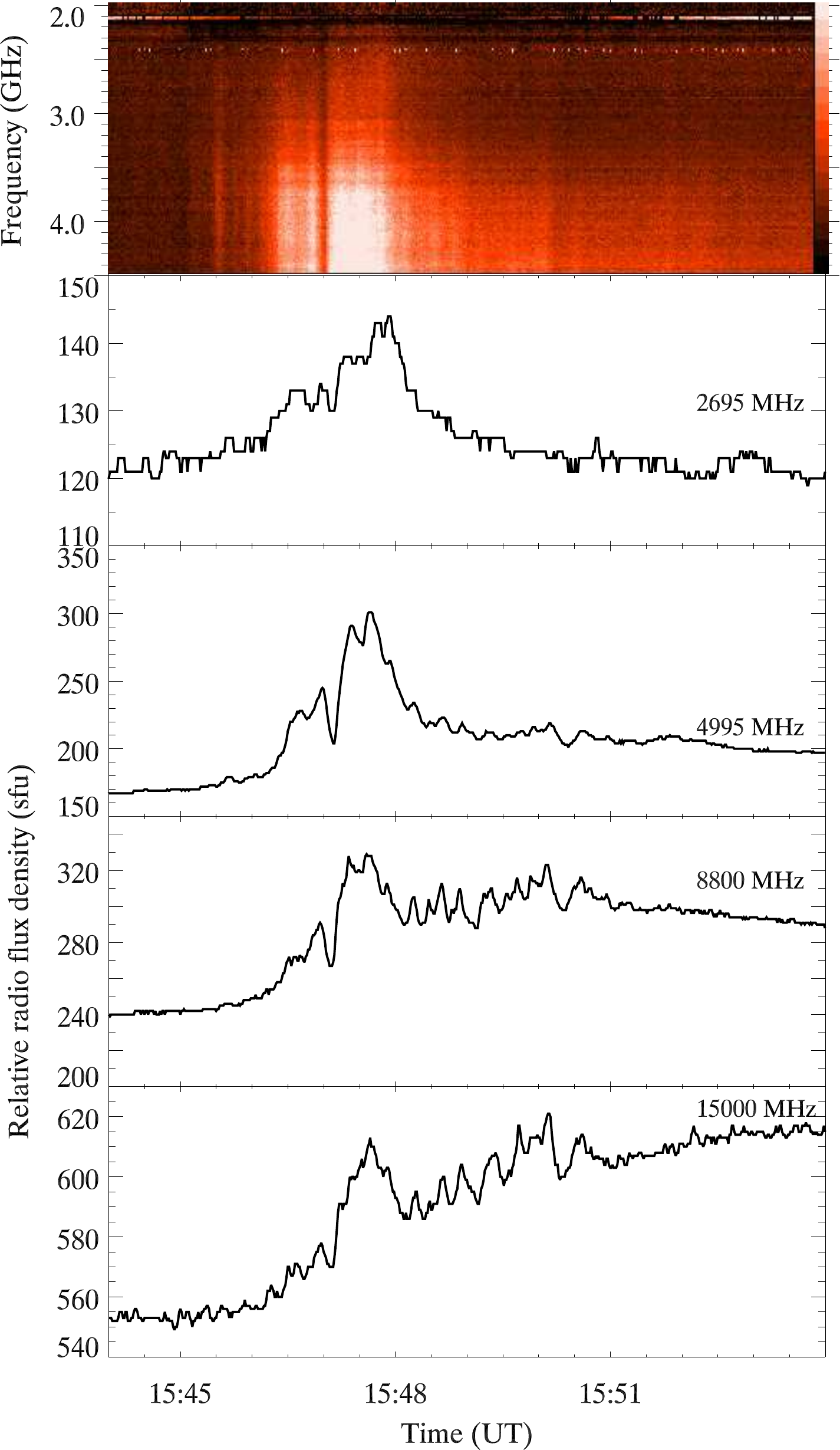}
}
\caption{Top panel: Ondrejov dynamic radio spectrum on 27 April, 2006 showing the
          intense DCIM radio burst during flare initiation.
         Additionally, there was no Type III burst during this time period
         (checked with Wind/WAVES spectrum). That means the opening of
         field lines did not take place during the flare energy release
         (i.e. during reconnection). The observed DCIM burst is the signature of particle acceleration from the
         reconnection site during loop-loop interaction/coalescence.
         Bottom panel: RSTN 1 sec cadence radio flux profiles in 2.6, 4.9, 8.8 and 15 GHz frequencies observed at Sagamore-Hill station.}
   \label{radio}
\end{figure}

We have used  Ondrejov dynamic radio spectrum data (2--4.5 GHz) during the flare \citep{jir1993,jir2008}.  This radiospectrograph uses a 3-m dish and 
wide band horn antenna as primary feed. The time resolution is 10 ms and the frequency band
is divided into 256 channels, which mean the frequency resolution is of about 10 MHz. Figure \ref{radio} (upper panel) displays the Ondrejov dynamic 
radio spectrum on 27 April, 2006 showing the intense DCIM radio burst during
flare initiation. Moreover, there was no Type III burst during this time period (checked
with Wind/WAVES spectrum). That means the opening of field lines did not take place
during the flare energy release (i.e. during reconnection).
%%%   %%%   %%%   %%%   %%%   %%%   %%%   %%% 
%
%%% Fig. 8 %%%
The DCIM burst starts in $\sim$2.5--3 GHz  frequency and continues upto 4.5 GHz. 
This frequency range covers the typical range of heights corresponding to reconnection site. 
The burst starts at 15:46 UT and continues upto 15:49 UT for the duration of $\sim$3 minutes.
The observed DCIM bursts reveal the signature of particle acceleration from the reconnection site during loop-loop
interaction/coalescence.

The US Air Force operates four solar radio observatories at various
locations around the world. These are collectively known as the Radio Solar Telescope Network or RSTN.
Each observatory monitors solar radio emissions on 8 discrete fixed
frequencies (245, 410, 610, 1415, 2695, 4995, 8800 and 15400~MHz) as well
as low frequency spectral emissions in the VHF band.
We have used the radio flux data (1 sec cadence) from Saga\-more Hill.
 We have selected four radio frequency bands of 2695, 4995, 8800 and 15000~MHz, which show significant
variations in the flux profiles. The radio burst is observed during $\sim$15:46--15:49 UT (Figure \ref{radio}, lower panels).  
The radio flux profiles in 4900 and 8800 MHz show double peak structures
associated with the coalescence of the loop-systems. 
It may be noted that second double peak structure is stronger in 
comparison to the first one, which shows that the superthermal electrons accelerated from a higher amount of pre-accelerated electrons
generated the last double peak \citep{karl2003}. 
After this burst, we observe the quasi-periodic oscillations specially in 4995, 8800 and 15400~MHz frequencies during 
$\sim$15:48--15:51 UT for the duration of $\sim$3 minutes, which may be attributed to modulations by 
MHD oscillations or nonlinear relaxational oscillations of wave particle interactions. 
Therefore, MHD waves can modulate the emissions from the trapped electrons  \citep{asc2004}.

The absence of Type III radio burst suggests the absence of opening of field lines during the reconnection process. 
Further, we do not see plasmoid ejection in soft X-ray images from the reconnection site. 
Therefore, the DCIM radio burst can not be interpreted as ejected plasmoid from the reconnection site.  
It should be noted that the burst starting frequency is $\sim$ 2.5-3 GHz, which corresponds to the typical
 height of post flare loops and originates in magnetic reconnection regions 
(i.e. plasma density of $\sim$ 10$^{10}$-10$^{11}$ cm$^{-3}$ ) \citep{asc2004}.  
As this burst continuation can be seen upto 4.5 GHz in the radio spectrum and further in 
 single frequencies radio flux profiles ( i.e. in 2.6, 4.9, 8.8 and 15 GHz). Therefore, we interpret 
these emissions due to nonthermal electrons accelerated from the reconnection site along the soft X-ray 
loop systems. This may be confirmed by the soft X-ray image at  15:47:02 UT, which shows the four footpoints due to precipitated electrons during the time of  radio burst.

\begin{figure}
\centering
{
\includegraphics[width=7cm]{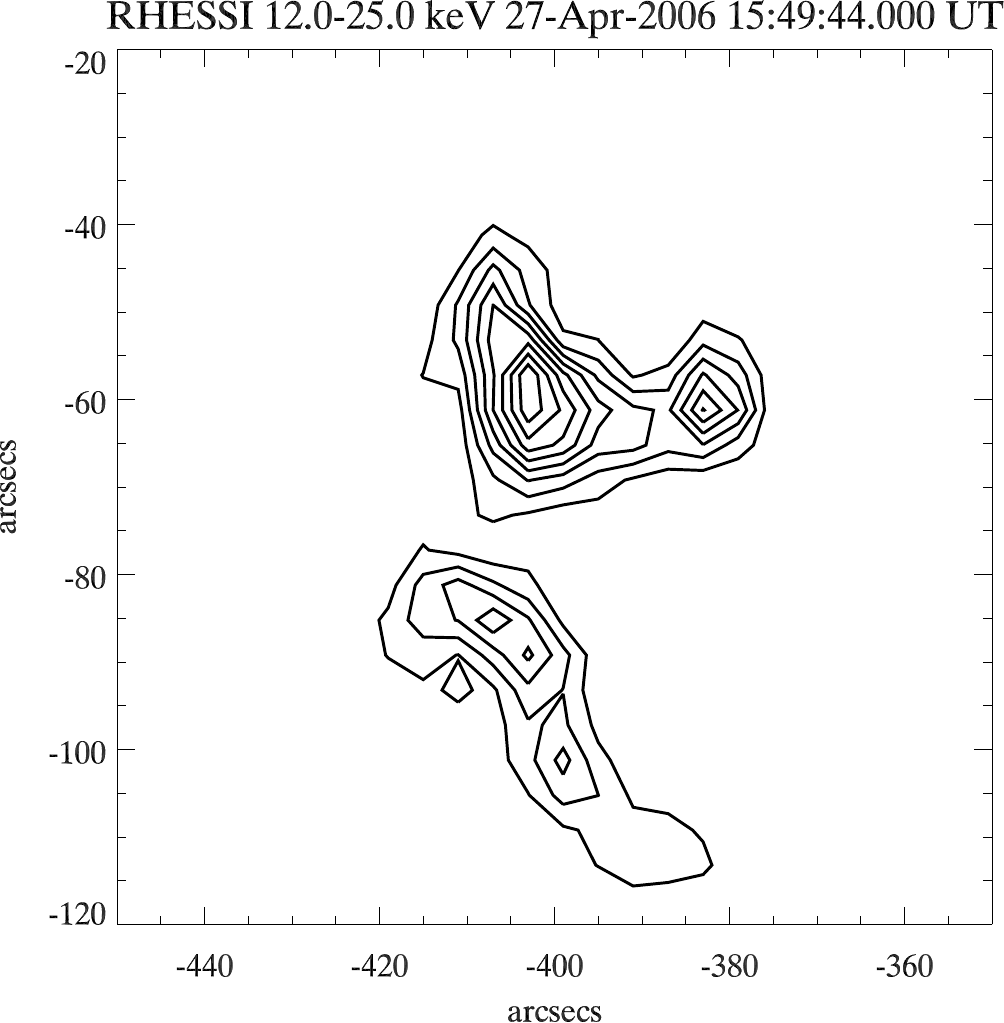}
\includegraphics[width=7cm]{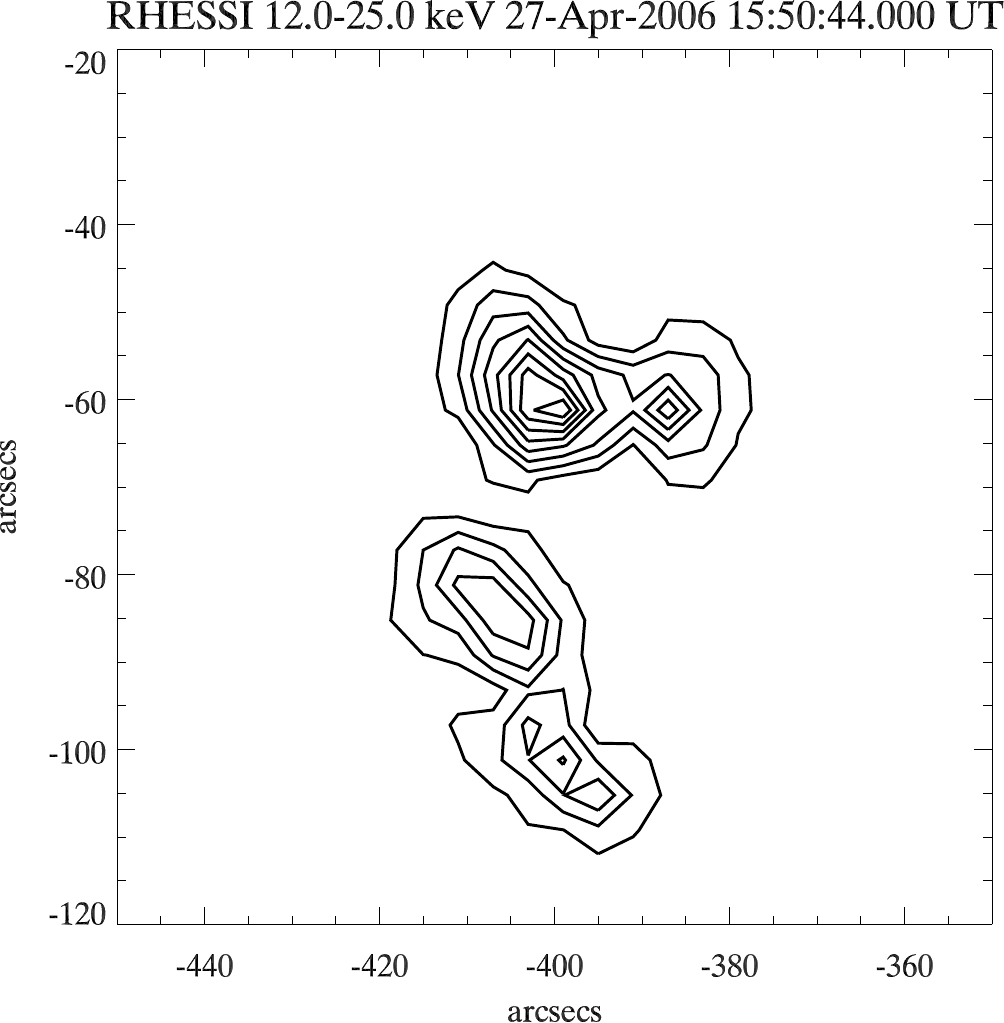}
}
\centering
{
\includegraphics[width=7cm]{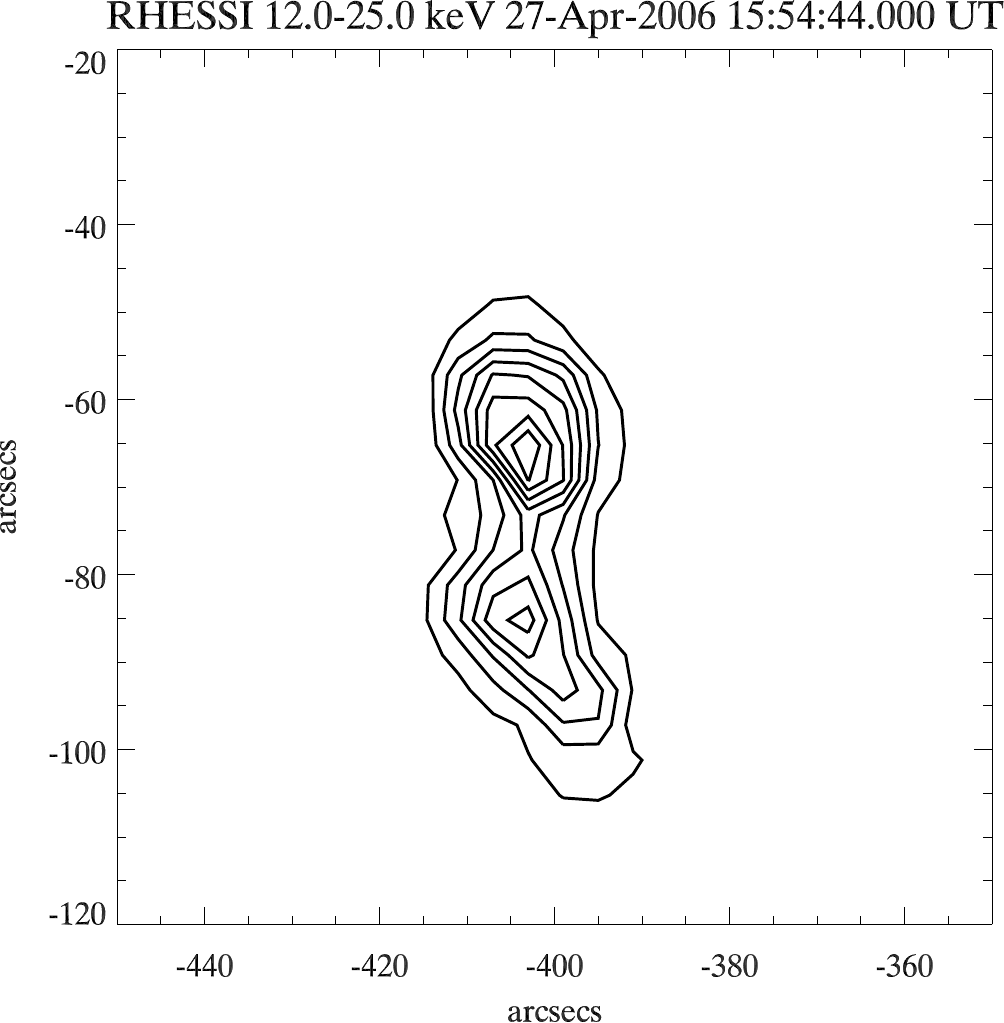}
\includegraphics[width=7cm]{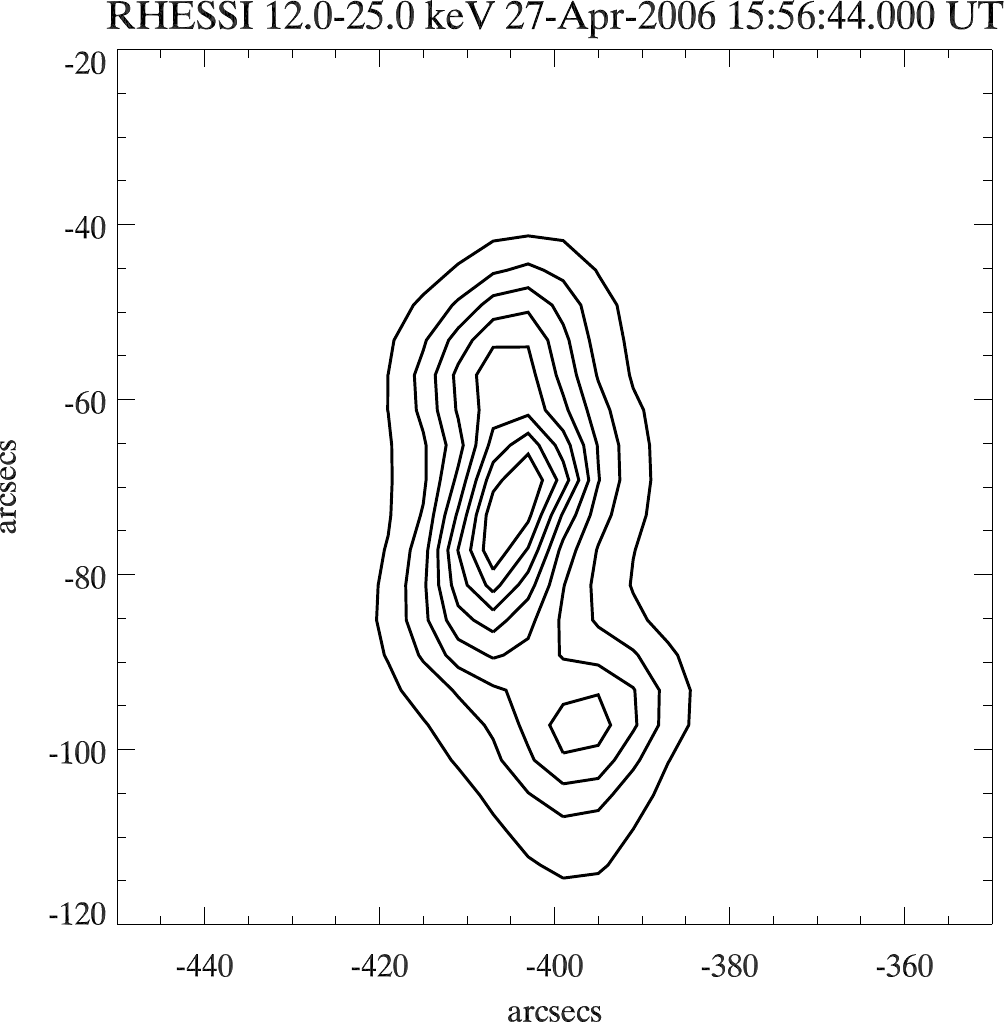}
}
\caption{RHESSI images in 12-25 keV energy bands reconstructed with the
         PIXON algorithm (contour levels for each image are 40$\%$, 60$\%$,
         80$\%$ and 95$\%$ of peak flux).}
\label{hessi1}
\end{figure}

\begin{figure}
\centering
{
\includegraphics[width=5cm]{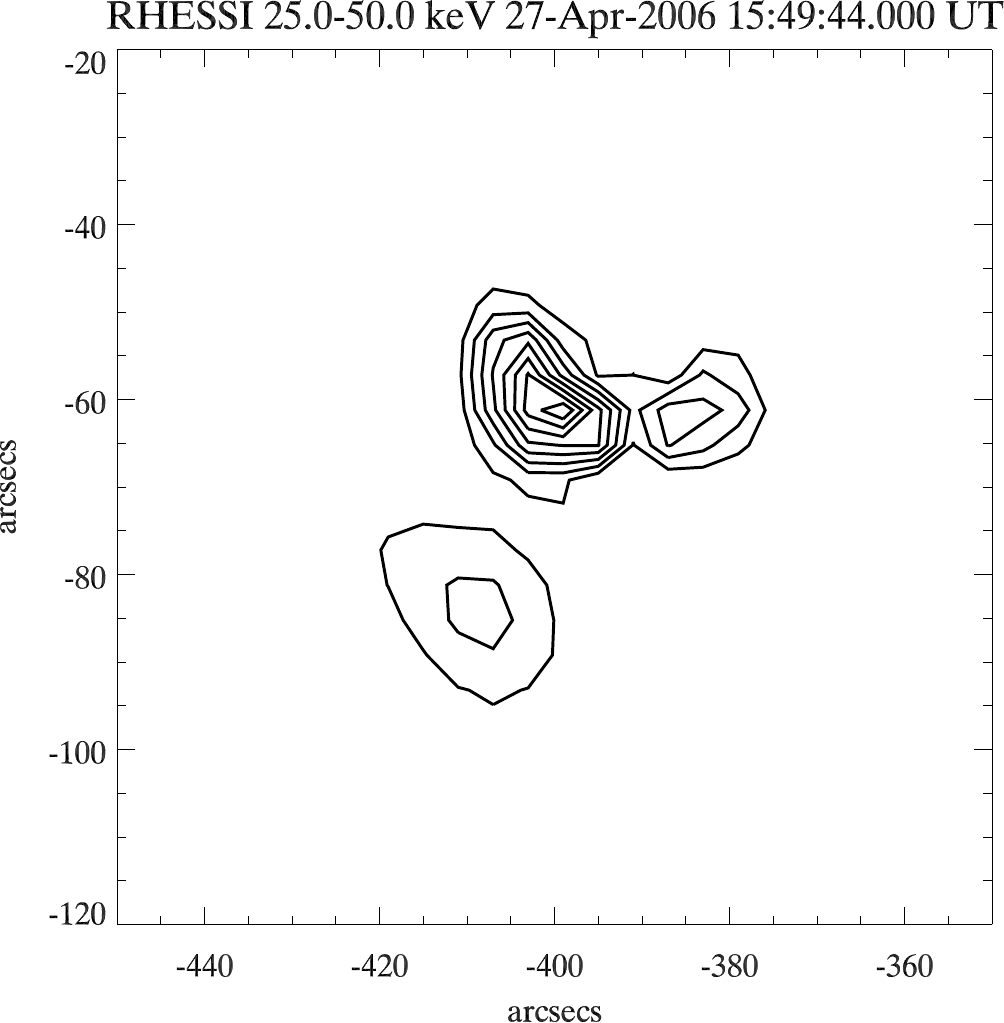}
\includegraphics[width=5cm]{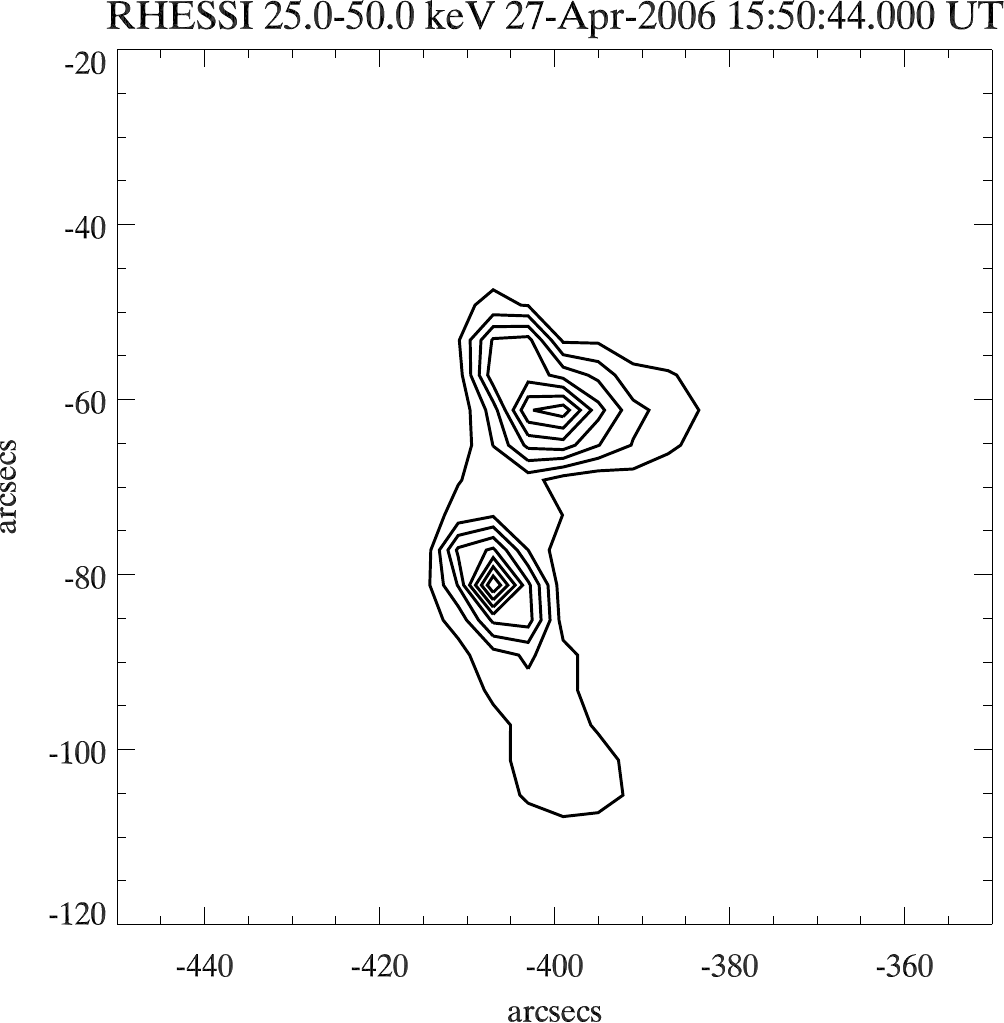}
\includegraphics[width=5cm]{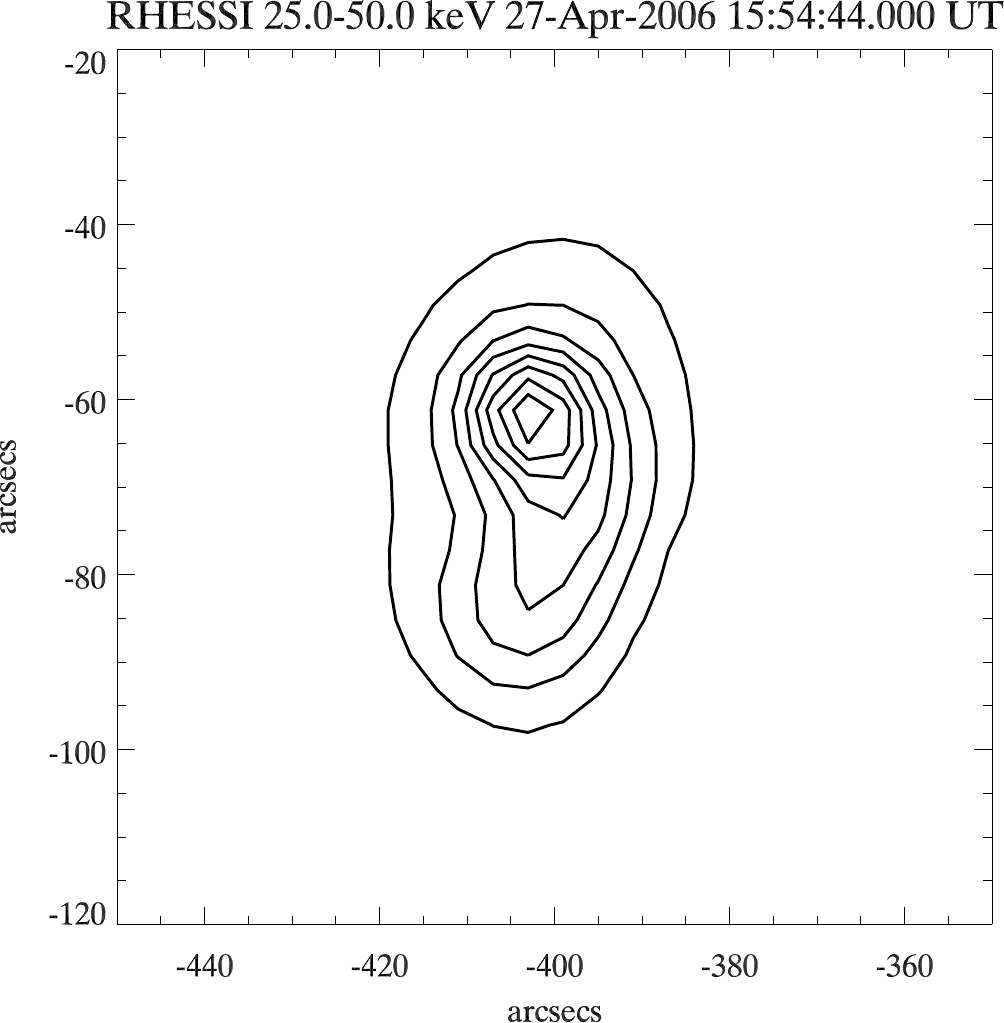}
}
\caption{RHESSI images in 25-50 keV energy bands reconstructed with the
         PIXON algorithm (contour levels for each image are 40$\%$, 60$\%$,
         80$\%$ and 95$\%$ of peak flux).}
\label{hessi2}
\end{figure}

The evolution of hard X-ray sources in two selected energy bands
(12-25 and 25-50~keV) of RHESSI instrument is shown in 
Figure \ref{hessi1} and \ref{hessi2}.     %%%   %%%   %%%   %%%   %%%   %%%   %%%   %%%   %%%
%
%%% Figures 9 and 10 %%%
%
These images have been reconstructed using PIXON method.
In both the energy bands, the two separated loop-top sources are
visible at 15:49 and 15:50~UT and then their coalescence resulting into
a single source (at 15:54 and 15:56~UT).
These images also provide the evidence of two loops coalescence.

%
%%%%%%%%%%%%%%%%%%%%%%%%%%%%%%%%%%%%%%%%%%%%%%%%%%%%%%%%%%%%%%%%%%%%%%%%%%%%%%
%
%
\subsection{Evolution of Active Region}
\begin{figure}
{
\includegraphics[width=5.27cm]{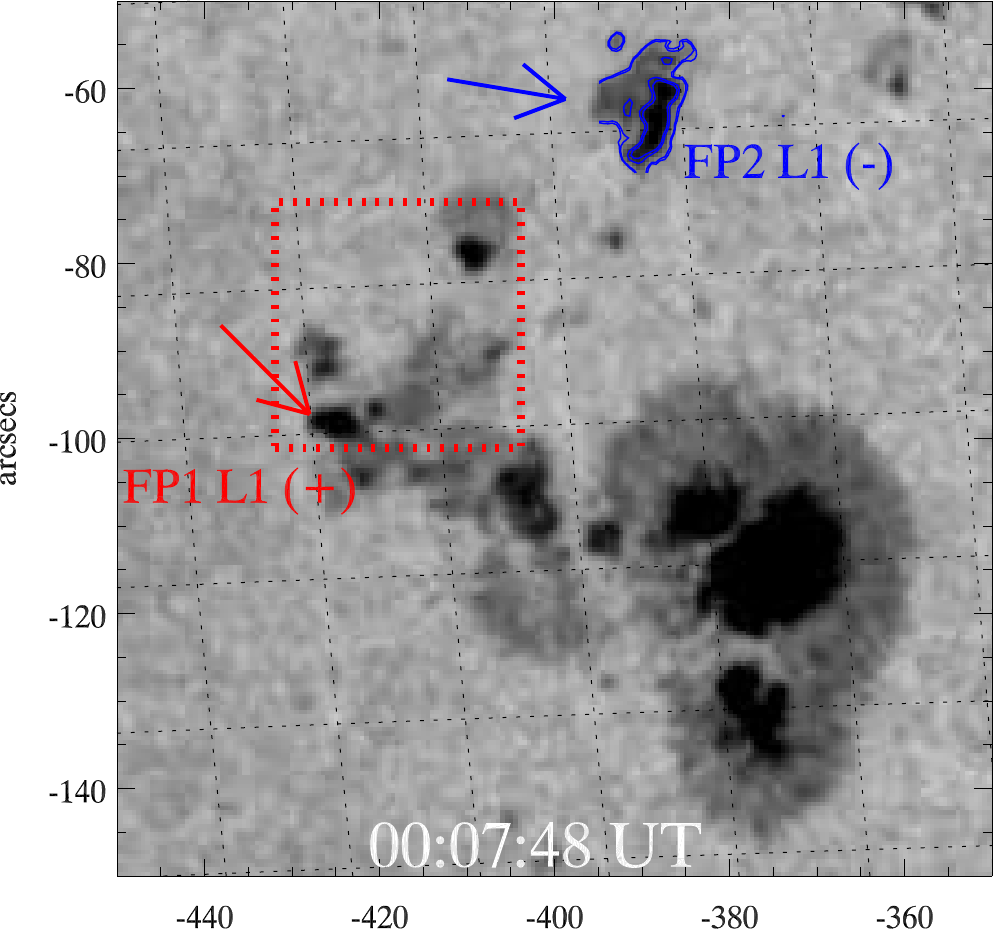}
\includegraphics[width=5cm]{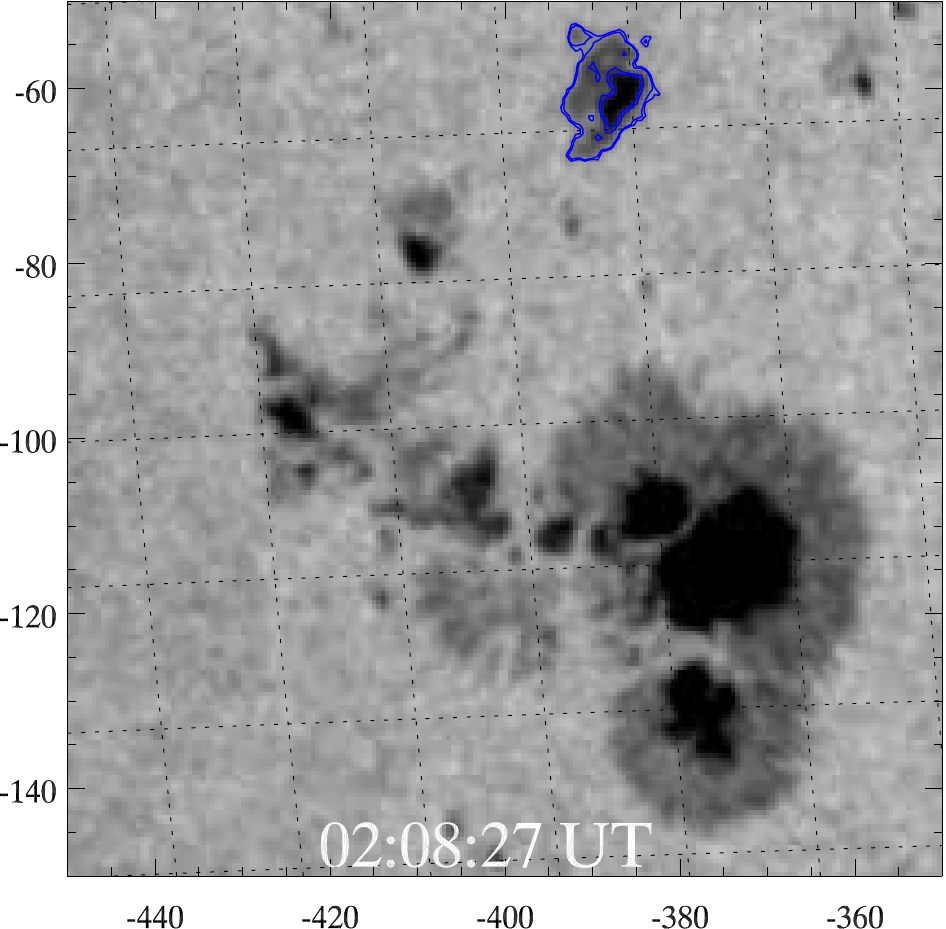}
\includegraphics[width=5cm]{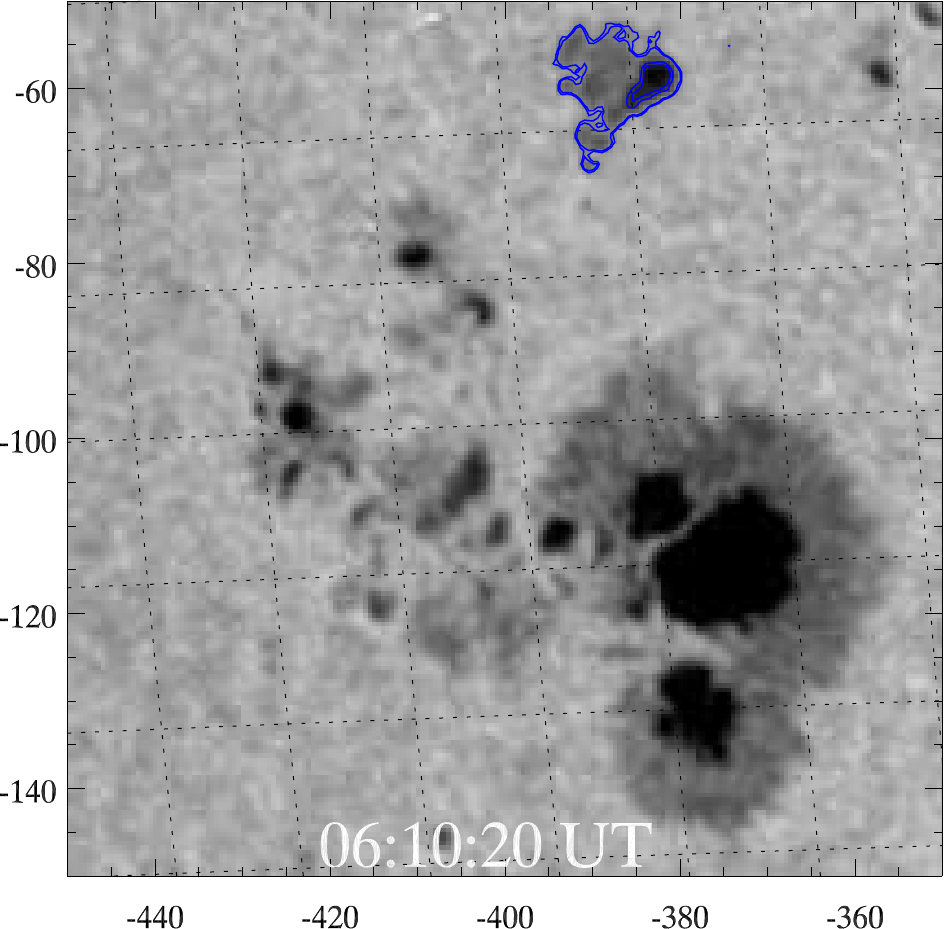}

\includegraphics[width=5.3cm]{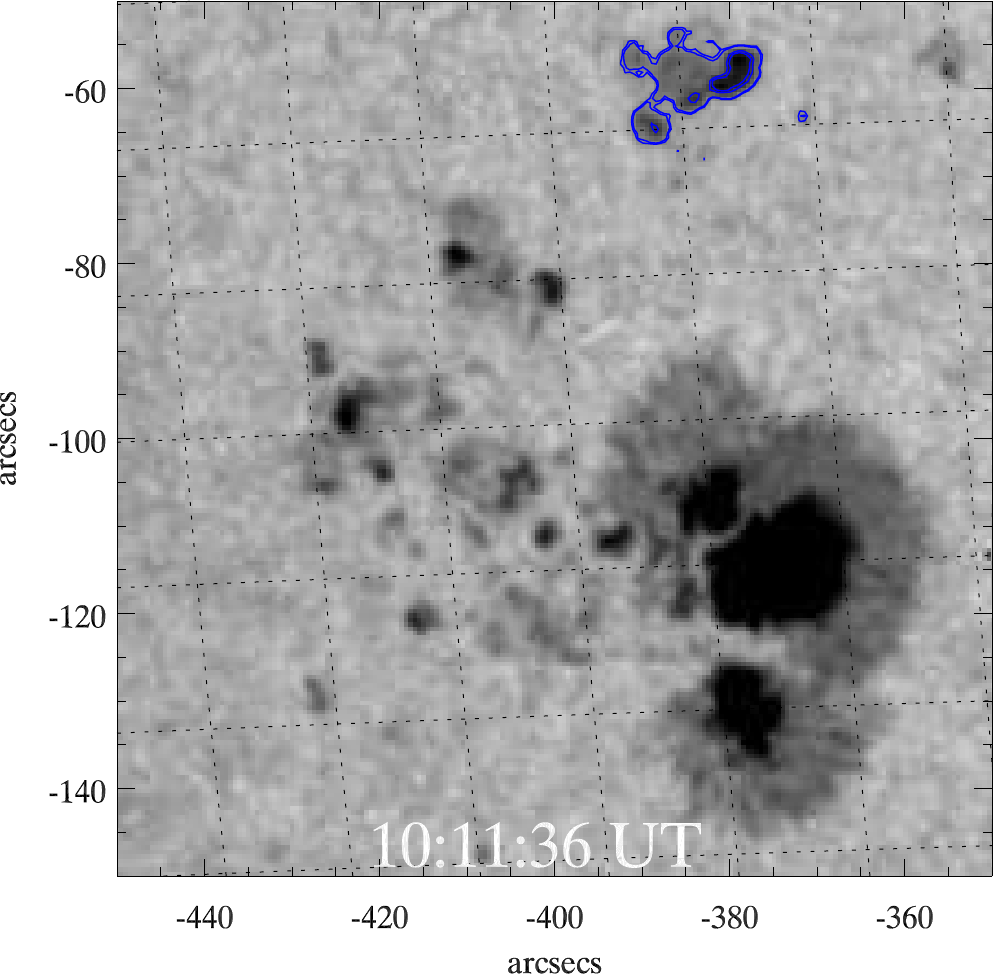}
\includegraphics[width=5cm]{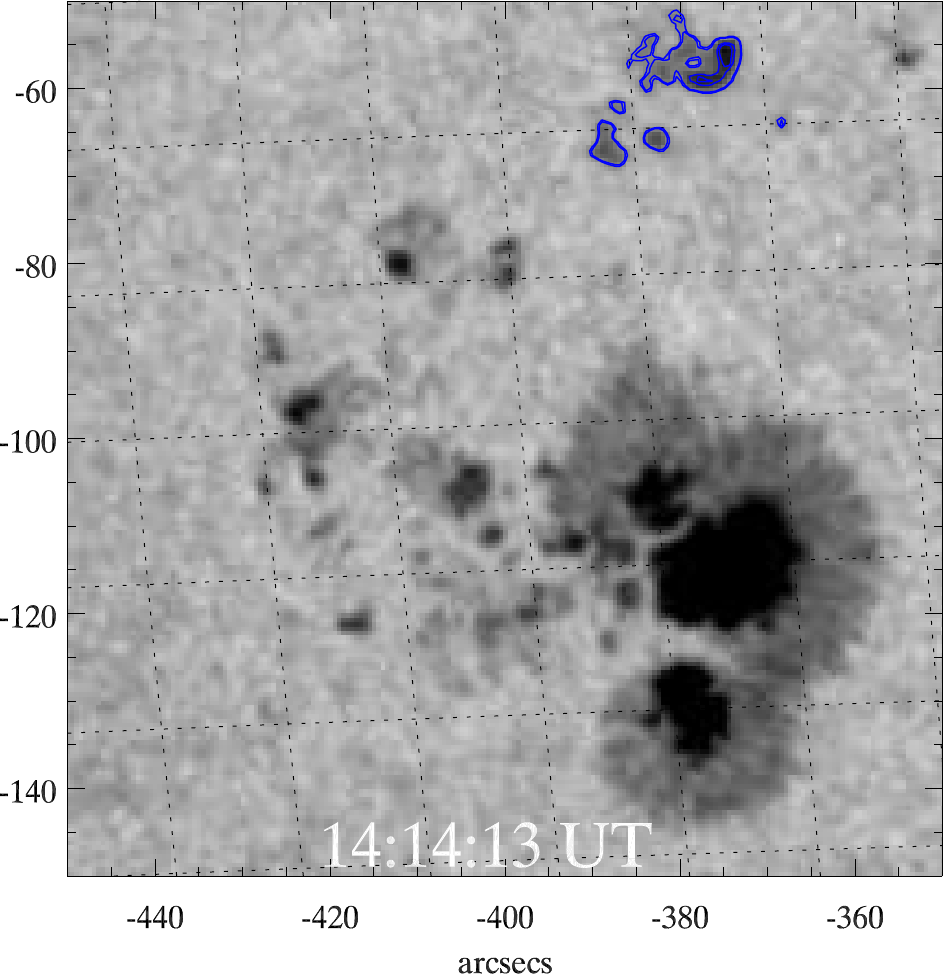}
\includegraphics[width=5cm]{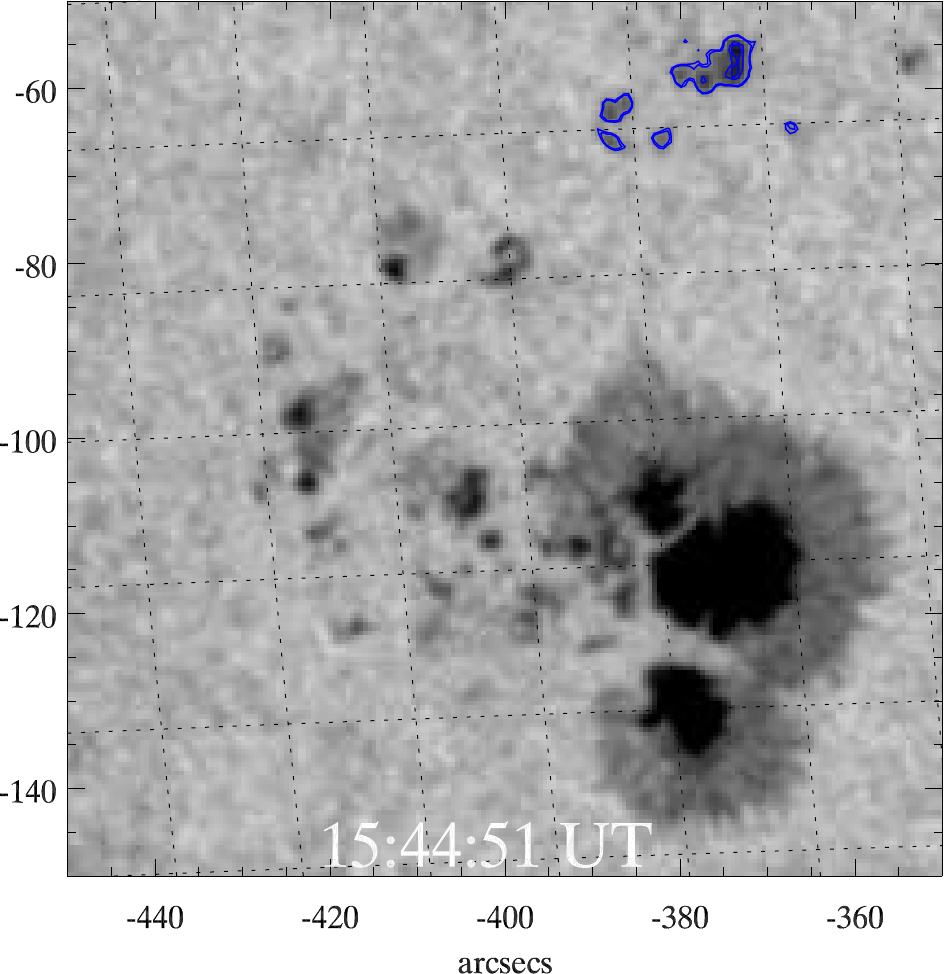}
}
\caption{ TRACE white-light images of the active region showing the linear/shear
         motion of negative polarity sunspot
         (indicated by blue contours). FP1 (red) and FP2 (blue) in the top first image show the `+ve' and `--ve' footpoints (indicated by arrows) of the lower loop system respectively.}
\label{tr_wl}
\end{figure}

\begin{figure}
\centering{
\includegraphics[width=6cm]{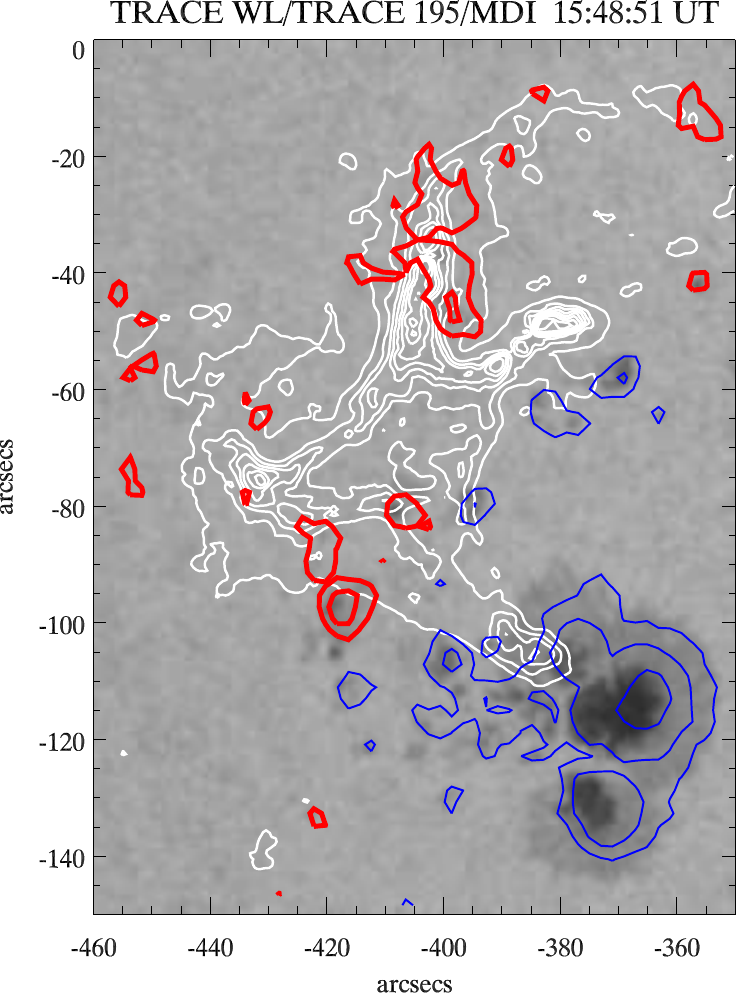}
\includegraphics[width=9.8cm]{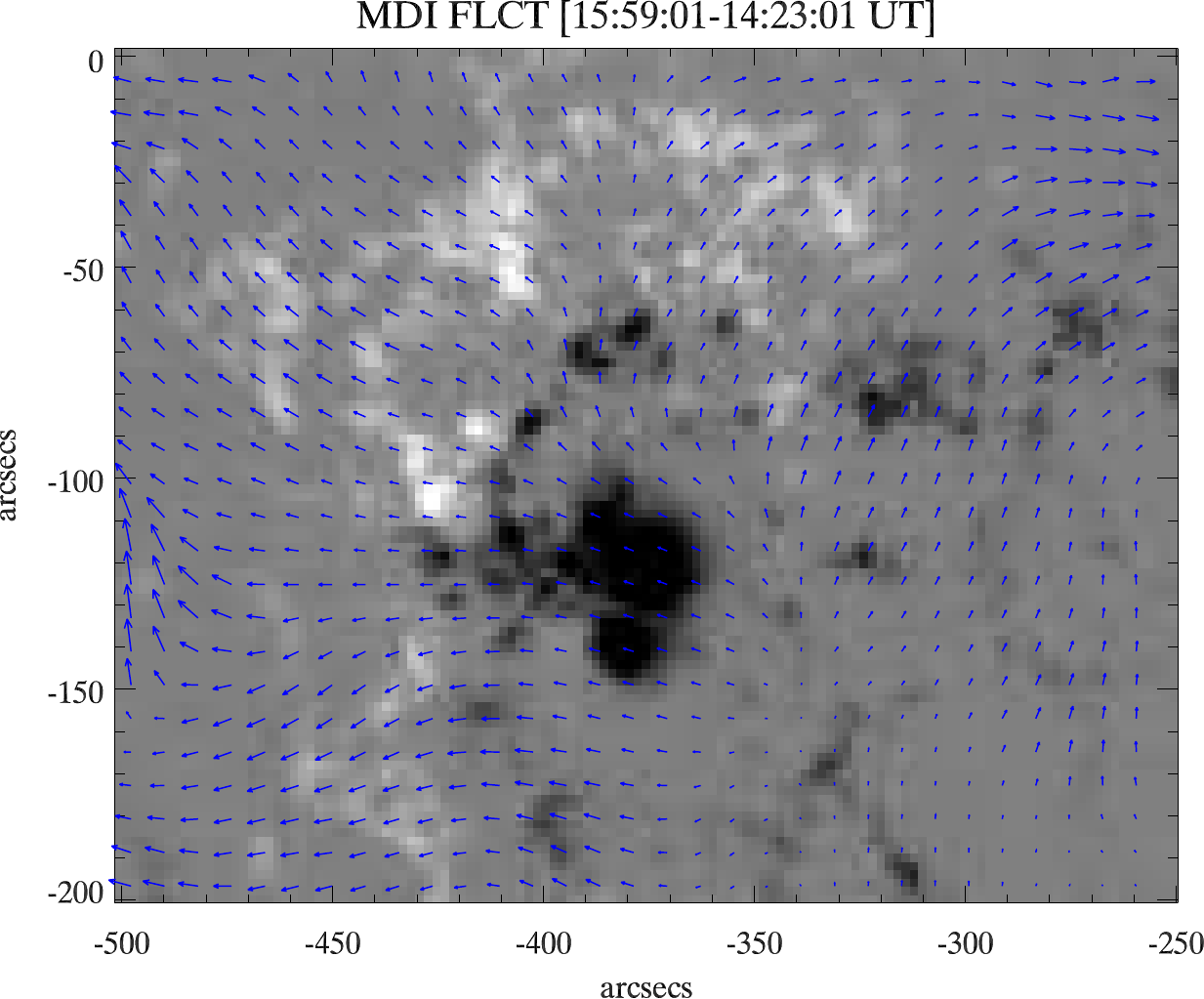}
}
\caption{Left: TRACE 195 \AA \ (white) and MDI magnetogram contours
         overlaid on TRACE white-light image. Red contours indicate the
         positive polarity sunspots whereas blue one show the negative
         polarity spots. The contour levels are $\pm 500, 
\pm 1000, \pm 2000, \pm 3000$ G. Right: The photospheric velocity map obtained
         from FLCT (Fourier Local Correlation Tracking) technique using
         SOHO/MDI magnetograms. The longest arrow corresponds to velocity
         of 0.291 km s$^{-1}$.}
\label{tr_wl_flow}
\end{figure}

\begin{figure}
\centerline{
\includegraphics[width=12cm]{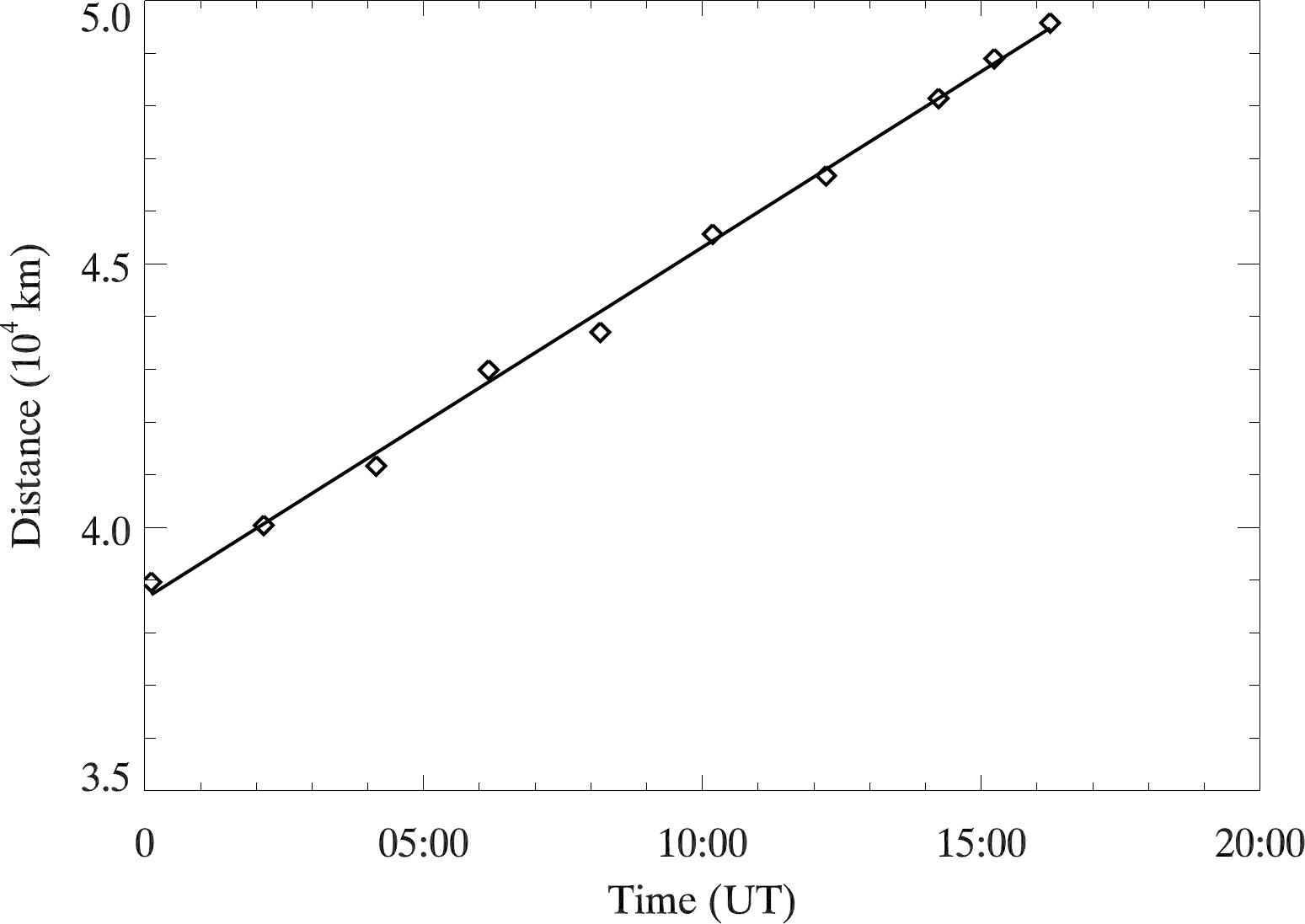}
}
\caption{The linear motion of negative polarity sunspot on
         27 April, 2006. One footpoint of the loop-system was anchored
         in this sunspot. The estimated speed of the sunspot from the
         linear fit is $\sim$ 0.2 km s$^{-1}$ (662 km h$^{-1}$).
         This motion probably caused the destabilization and interaction
         in the loop systems.}
\label{shear}
\end{figure}

Figure \ref{tr_wl}
%
%%% Fig. ? %%%
%
displays the selected images of TRACE white-light of active region on 27
April, 2006.  FP1 (red) and FP2 (blue) in the top first image show the `+ve' and `--ve' footpoints (indicated by arrows) of the lower loop system respectively.
The careful investigation of the TRACE movie reveals the linear/shear motion of
small sunspot of negative polarity (indicated by blue contours) across the
neutral line.
We have made the time-distance plot to quantify the linear translational motion of the
sunspot.
 From the linear fit to the data points, the speed of this motion is estimated as 
$\sim$0.2 km s$^{-1}$ (662 km~h$^{-1}$) (see Figure \ref{shear}).
To identify the foot\-point of the related loop-system anchored in this spot,
we  overlaid MDI and TRACE 195 \AA \ contours over the white-light image
(refer to Figure \ref{tr_wl_flow}, left).
%
%%% Fig.
%
This image reveals that one foot\-point of the loop-system is anchored in
this spot. In order to view the photospheric horizontal flow pattern in and around the
active region, we use the Fourier Local Correlation Tracking Technique (FLCT)
on SOHO/MDI images. The FLCT method is described by \citet{fisher2008}.
The main input parameters for this technique are, two images f1
and f2, the pixel separation scale ($\Delta$s) and time separation ($\Delta$t), and a Gaussian
window size scale ($\sigma$). This routine calculates the velocity (2D) by maximizing
the cross-correlation of each image when weighted by the Gaussian window centered 
on each pixel location. In our study, we use the two SOHO/MDI
frames at different times before the flare. After a careful investigation,
a Gaussian window with a standard deviation of 15$^{\prime\prime}$ was chosen.
The right panel of Figure \ref{tr_wl_flow} displays the photospheric velocity map obtained from
FLCT technique using SOHO/MDI magnetograms. 
The longest arrow corresponds to velocity of 0.291 km s$^{-1}$. 
It may be noted from the flow map that the small, negative polarity spot shows the 
clockwise shear flow motion whereas the positive polarity region (in which another footpoint was anchored 
of the lower loop-system) shows counter-clockwise flow motion.
This linear translational motion as evident in TRACE white light images as well as velocity shear flows as evident in FLCT 
images near the spots most likely indicate the triggering of the shear in their locations. This 
physical mechanism most likely plays a role in the energy build-up for flare
and generates the coalescence instability in the lower loop-system.

%
%
%%% Sub-section 2.3 %%%%%%%%%%%%%%%%%%%%%%%%%%%%%%%%%%%%%%%%%%%%%%%%%%%%%%%%%%
%
\subsection{Magnetic Topology of the Interacting Loop-Systems}
   \label{sub:topology}
\begin{figure}
\centerline{
\includegraphics[width=0.7\textwidth]{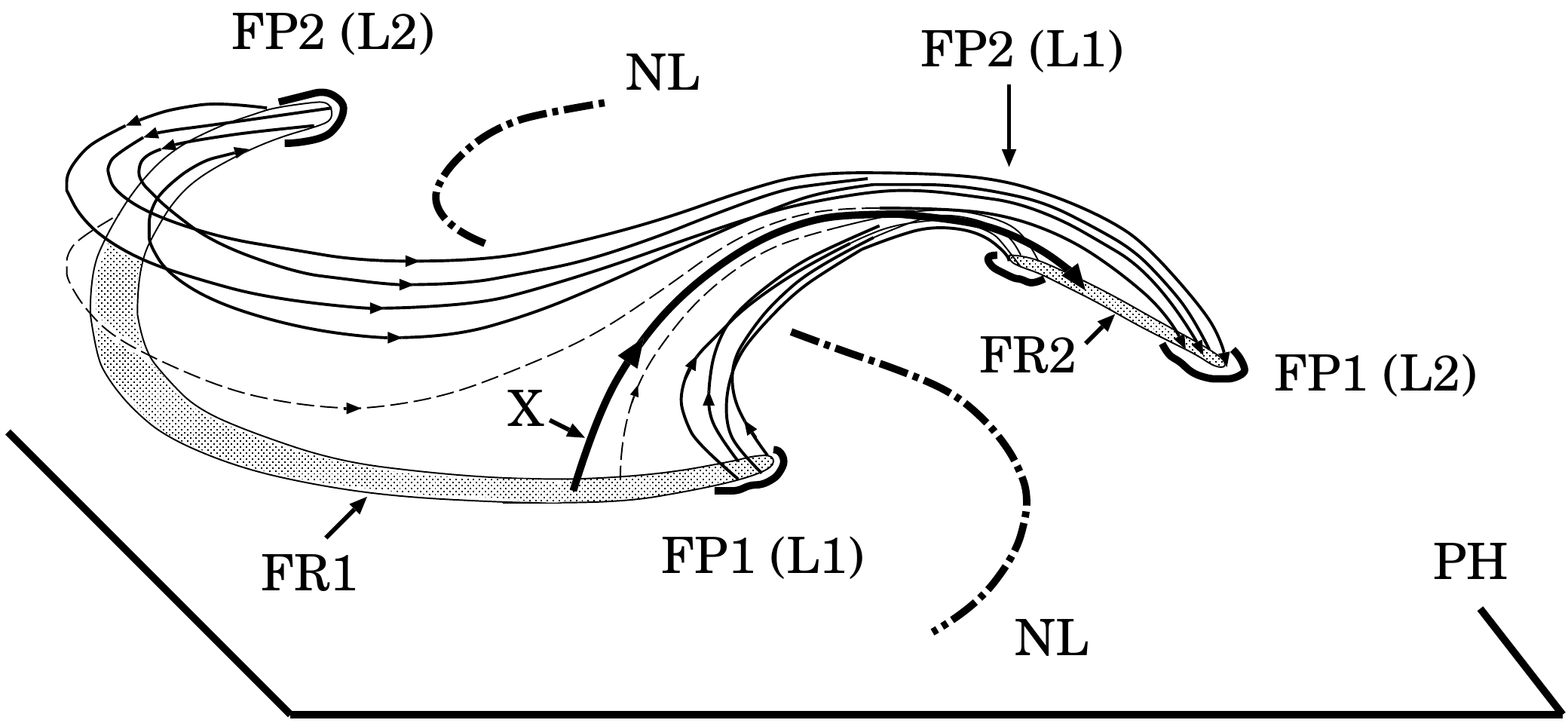}
}
\caption{Magnetic field lines that connect the H$ \alpha $ kernels
         kernels FP1 (L1), FP2 (L1), FP1 (L2), and FP2 (L2) are passing
         through a region of primary energy release located somewhere
         near the top of the separator X.
         The flare ribbons FR1 and FR2 are formed where these field lines
         cross the photospheric plane PH.
         NL is the neutral line of photospheric magnetic field.
         Chromospheric evaporation creates a picture of the crossing
         soft X-ray loops.}
\label{topology}
\end{figure}

 In this Section, we discuss the large-scale structure of a
magnetic field responsible for the solar flare.
The soft X-ray image of the flare clearly reveals the two large
solar loops (L1 and L2) crossing to each other and exhibit the X-type interaction.
The chromospheric images (H$\alpha$ and TRACE 1600~\AA) show
the two ribbon morphology with the four kernels, i.e. four
foot\-points of the reconnected loops.
We illustrate these features of the interacting loop-systems in
terms of the {\em topological\/} models (see ch.~3 in
\citet{somov2007}).
Figure \ref{topology}
%
%%% Fig. 14 %%% NEW FIGURE 14 %%%
%
displays the field lines that connect the H$ \alpha $ kernels:
FP1 (L1) with FP2 (L1), and FP1 (L2) with FP2 (L2).
The shadowed regions FR1 and FR2 indicate the flare ribbons.
They are located on both sides of the photospheric neutral line
NL.
Chromospheric evaporation along the reconnected field lines
creates the SXR loops that look like they are crossing or
touching each other somewhere near the top of a magnetic-field
separator X.
The loops and ribbon morphology shown in the observations
qualitatively matches with this cartoon.

It is very likely that, in addition to what is shown in
Figure \ref{topology},
%
%%% Fig. 14 %%%
%
the electric currents and twisted magnetic fields can be created
inside the interacting loops by some under-photospheric or
photospheric mechanism observed in the photosphere as shear
motions or rotations.
Such currents certainly must exist in complex active regions
with sunspot rotation and large-scale photospheric shear flows.
If the currents are mostly parallel, they attract each other and
can give energy to a flare
\citep{gold1960}.
On the other hand, according to the simplified topological model
presented in Figure \ref{topology},
%
%%% Fig. 14 %%%
%
the flare energy comes from an interaction of magnetic fluxes
that can be mostly potential.
If this would be the case, the flare energy should be stored
before a flare mainly in slowly-reconnecting current layer at
the separator of coronal magnetic field.
This possibility seems to be in agreement with the quad\-ru\-pole
reconnection model of the solar flares.
The morphology of the loops is also in agreement with the PFSS
extrapolation of photospheric magnetic fields into the corona.
Therefore, we consider both the models firstly from the view-point
of global magnetic configuration of a quad\-ru\-pole-type active
region taking into account the interacting electric currents.

\begin{figure}
\centerline{
\includegraphics[width=0.65\textwidth]{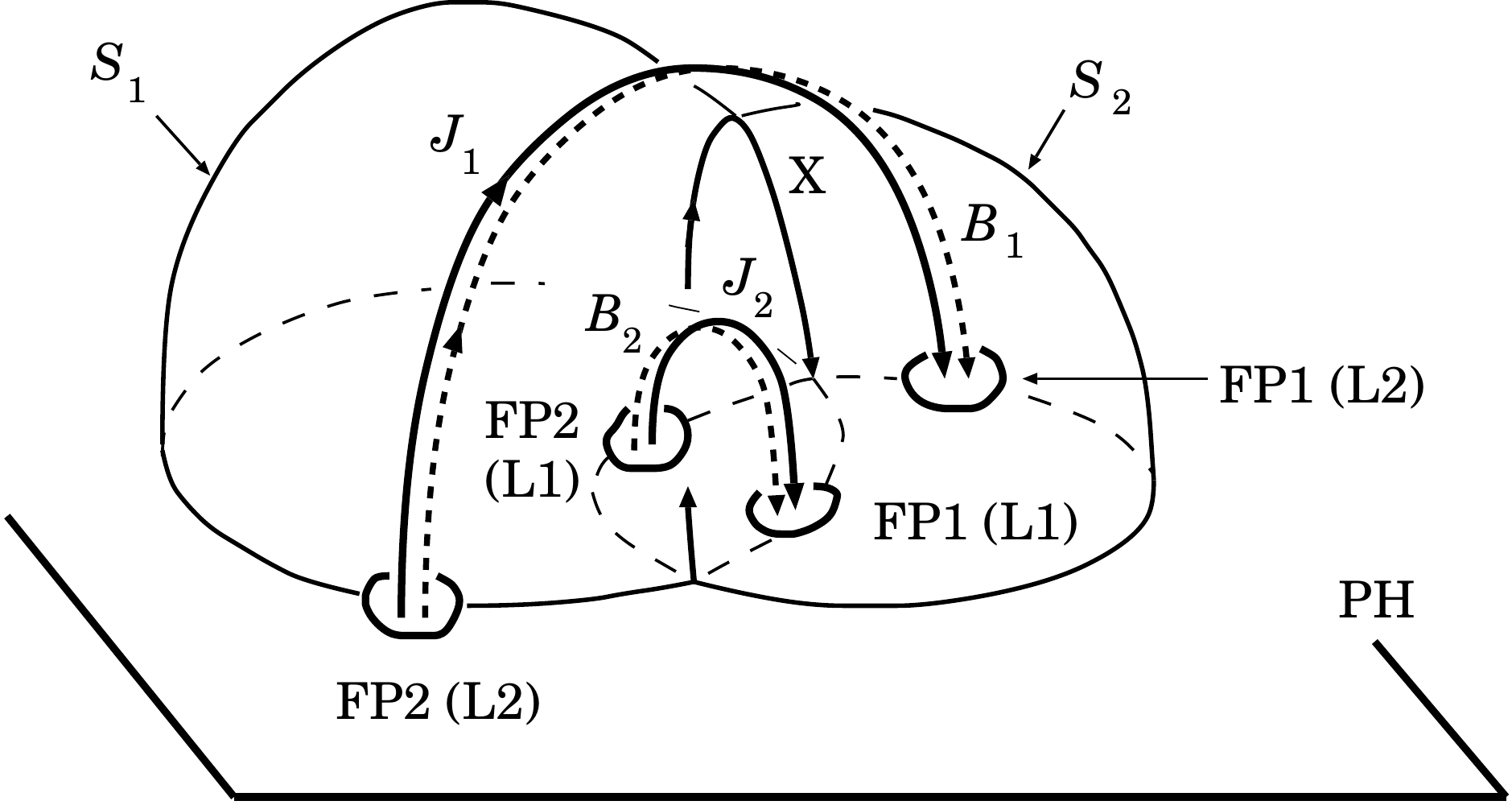}
}
\caption{ A 3D model of the coronal magnetic field
         with two interacting electric currents $ J_{1} $ and $ J_{2} $.
         Four magnetic fluxes of different linkage are separated by
         the separatices $ S_{1} $ and $ S_{2} $ that cross at the
         separator X above the photospheric plane PH.
         The two field lines~$ B_{1} $ and $ B_{2} $ connect the kernel
         FP2 (L2) with FP1 (L2) and the kernel FP2 (L1) with FP1 (L1).
         The coronal currents are distributed somehow inside the two
         magnetic cells and are shown schematically as the total
         currents~$ J_{1} $ and $ J_{2} $ along the field lines~$ B_{1} $
         and $ B_{2} $.}
\label{currents}
\end{figure}

Figure~\ref{currents}
%
%%%%%% Fig. 15 NEW %%%%%%   %%%%%%%%%%%%%%%%%%%%%%%%%%%%%%%%%%%%%
%
illustrates the {possible configuration of two large scale} coronal
currents~$ J_{1} $ and $ J_{2} $ distributed inside two different
magnetic cells, i.e. the two magnetic fluxes of different
linkage that interact and reconnect at the separator~$ X $.
The two field lines~$ B_{1} $ and $ B_{2} $ belong to the
magnetic cells that connect the kernel FP2 (L2) with FP1 (L2)
and the kernel FP2 (L1) with FP1 (L1) respectively.
The coronal currents are distributed somehow inside the two
different magnetic cells and shown schematically the total
currents~$ J_{1} $ and $ J_{2} $ along the field lines~$ B_{1} $
and $ B_{2} $.

If the field lines~$ B_{1} $ and $ B_{2} $ near the current layer
along the separator have an opposite direction component, then they
can be reconnected.
If the two current systems~$ J_{1} $ and $ J_{2} $ flow more or
less in the same direction, then they also attract each other according to
\citet{gold1960}.
The components of the magnetic field transversal to the separator
reconnect together with electric currents flowing along them
\citep{hen1987,somov1992}.
In this way, with a perpendicular magnetic field inside the place
of interruption, magnetic reconnection can create local
interruptions of the electric currents in the solar atmosphere.
If these currents are highly concentrated, their interruption can
give rise to strong electric fields that accelerate the energetic particles and
can contribute significantly to the flare energetics.

What factors do determine the rate of magnetic reconnection in
the current layer at the separator? --
Let us consider the magnetic fields created by the
currents~$ J_{1} $ and $ J_{2} $.
These additional or secondary fields play the role of the
longitudinal magnetic field near the reconnecting current layer.
Being superimposed on the large-scale potential field, they
create the two types of field line spirals, i.e., left-handed and
right-handed.
When looking along the positive direction of the field
lines~$ B_{1} $ and $ B_{2} $, we see the two opposite
orientations for the spirals namely to the right for the
{\em dextral\/} structure and to the left for the
{\em sinistral\/} one.
Depending on this handedness property known as
a {\em chirality\/} also that depends on the angle between the
currents~$ J_{1} $ and $ J_{2} $, magnetic reconnection of
electric currents will proceed faster or slower
\citep{hen1987}.

As evident in the observations as well as in the theoretical baseline,
the X-type reconnection may produce the plasma jets.
However, we have no observational signature of such jets in our
observations.
In the flare under consideration, the reconnected fast outflows
from a current layer relax quickly because they interact with
(i) closed field lines of a quadru\-pole-type of the active
region (recall that there was no type III radio\-burst, thus
the opening of field lines did not take place during the flare
energy release, i.e. reconnection);
(ii) chromospheric evaporation upflows (the energy released in
closed magnetic configuration goes into impulsive heating of the
upper chromosphere to high temperatures that is why the soft
X-ray images become so bright quickly).

%%%%%%%%%%%%%%%%%%%%%%%%%%%%%%%%%%%%%%%%%%%%%%%%%%%%%%%%%%%%%%%%%%%%%%%%%%%%%%

%%%%%% Section 3 %%%%%%   %%%%%%%%%%%%%%%%%%%%%%%%%%%%%%%%%%%%%%%%%%%%%%%%%%%%
%*****************************************************************************

\section{SOME THEORETICAL ESTIMATIONS}

The RHESSI temporal images (12-25 and 25-50 keV) reveal the coalescence of
the loop-top sources of the interacting loop system.
The two loop-top sources merge approximately vertical  in the RHESSI field
of view.
Therefore, the lower bound change of the distance of the two approaching
loops is
% (1)
\begin{equation}
    \Delta l_{coal} \approx 22000\, \, \, {\rm km} \;
\end{equation}
and the elapses time is
% (2)
\begin{equation}
    \Delta\tau_{coal} \approx 420 \, \, {\rm s} \;
\end{equation}
The coalescence instability may activate in the observed interacting loops
system, which is the effect that merges the two isolated magnetic islands
into a single one  \citep{har2001b,har2001a,asc2004}.
This type of instability evolves in two phases, i.e. First phase in pairing
of the current filament/loops as in ideal MHD process, while the second as
the resistive phase of pairwise reconnection between the approaching current
carrying flux tubes.
% It remains to look at their assumptions and relevancies %
The numerical MHD simulations reveal the different phases of coalescence
instability in ideal/resistive solar plasma \citep{sch1997}.

The characteristic time scale of the ideal phase of coalescence instability
is the multiple of Alfv\'enic transit time \citep{asc2004}:
% (3)
\begin{equation}
    \tau_{coal}=\frac{1}{q_{coal}}. \frac{l_{coal}}{v_A}, \, \,
\end{equation}
where
% (4)
\begin{equation}
    q_{coal}=\frac{u_{coal}}{v_A},
\end{equation}

The l$_{coal}$, u$_{coal}$ and v$_A$ are respectively the distance between
approaching loops, approaching velocity and local Alfv\'enic speed.
Using equation (3) and (4), the differential coalescence speed %%% :
% (5)
\begin{equation}
    \Delta u_{coal}=\frac {\Delta l_{coal}}{\Delta \tau_{coal}},
\end{equation}

Therefore, using the observationally estimated values as mentioned in
equation (1) and (2), we get the average coalescence speed as $\sim$52 km~s$^{-1}$.
TRACE 195~\AA \ images also show the interacting and paired loops.
Using these images, the projected distance-time profile of the interaction
region (i.e. converging motion at the interaction site) has been presented
in Figure 4.
%
%%% Fig. 4 %%%
%
The average converging speed of the interaction region is estimated as
$\sim$30 km s$^{-1}$. The approximate approaching velocity of one magnetic island
of a loop is evident as $\sim$26 km s$^{-1}$. The resemblance in these two speeds is in agreement with loop coalescence.

By assuming the typical Alfv\'enic speed at the interaction region as
$\sim$1000 km s$^{-1}$ and the projected distance between the approaching
loops ($\Delta$l$_{coal}$$\approx$22000 km), the estimated Alfv\'enic transit
time of the region will be $\sim$22 s.
Therefore the coalescence will occur $\sim$20 $\tau$$_A$ for our
observation, which is rather longer as predicted in various simulation
results explained by \citet{sakai1996} as well as \citet{tajima1982}
under various assumptions of the model atmosphere.
However, for L$\sim$62800 km, $\tau$$_A$=16 s, the Reynolds number (S=R)=500,
n$_e$=10$^{10}$ cm$^{-3}$ and B$_Z$=90 G, \citet{milano1999} have found
that two loops coalesces at t=11$\tau$$_A$ and the magnetic energy and
even its dissipation enhanced.
The loop coalescences time depends upon various atmospheric parameters,
and therefore further simulations will be interesting to study the dynamics
and energetics of our observed coalesced loops.

We can estimate the amount of energy ($ {\cal E}_c $)
%
%%% In order to distinguish from electric field %%%
%
available due to coalescence instability \citep{tajima1982,smartt1993} by:
% (6)
\begin{equation}
    {\cal E}_c \approx \frac{LB^2a^2}{2} \, \ln \frac{L}{a}
\end{equation}
where $ L , B $ and $ a $ are length of the reconnecting region,
loop magnetic field and radius of current loop respectively.
We take $ B \approx 100 $~G, $ L \approx 22000 $~km and
$ a \approx 11000 $~km, which gives
% (7)
\begin{equation}
    {\cal E}_c \approx 1.0\times10^{31} \, \, {\rm ergs} \, .
\end{equation}
Therefore, this value is comparable with the energy released during M-class
flare.
%
%%% Attention of the Authors !!! %%%   %%%%%%%%%%%%%%%%%%%%%%%%%%%%%%%%%%%%%%%
%

In general, the total magnetic field energy of the currents generated by
photospheric vortex flows, sunspot rotation or shear flows in the photosphere
can exceed the energy of even the largest flares.
However, in contrast to thin current layer at the separator, these currents
are typically dispersed over a large volume of magnetic flux tubes in
the corona.
The dissipation rate of the currents so distributed in the coronal plasma
of very high conductivity is vanishingly small.
However, their interaction with each other and with the current layers
at the separator is not small and must be treated within the framework of
the global electro\-dynamical coupling of a flare active region or complex.

%
%%% Attention of the Authors !!! %%%   %%%%%%%%%%%%%%%%%%%%%%%%%%%%%%%%%%%%%%%
%
As we saw in Section~\ref{sub:topology},
%
%%% Sub-section   %%%
%
a distinctive feature of this interaction is that the separator is orthogonal
(in the sense of the magnetic field topology) to both systems of electric
currents~$ J_{1} $ and $ J_{2} $.
For this reason, not only the magnetic field components associated with the
current layer, but also the longitudinal (guiding) components with respect
to the separator are reconnected.
Therefore, not only the energy associated with the current layer at the
separator, but also a part of the energy of the currents generated by the
photospheric vortex flows, sunspot rotation and shear flows is released in
the flares \citep{hen1987}, see also \citet{somov2002}.

 All the above have been concerned with the large-scale
structure of magnetic fields and electric currents in large
solar flares that can be qualitatively described in the main
features by the simplified topological models.
However, in actual flares there are many
different structures of different scales including the
smallest ones. In the flare under consideration, we see many interacting small
flux threads/tubes (e.g., Figure~\ref{tr_195}). Moreover, the image at 15:48 UT in this figure shows the
loop-loop interaction and formation of 'X' point in between the
interacting loop-system.
So, it is likely that the observed flare was caused by
interactions of not two but the multitude of the loops, forming
more-or-less parallel systems and visible in low-resolution
images as single wide loops.
From theoretical point of view, this presumably means that the
distributed currents $ J_{1} $ and $ J_{2} $ are deeply pinched
in many thin current filaments.
Therefore, we observe some average picture of reconnection with
some average reconnection rate.

%%%%%%%%%%%%%%%%%%%%%%%%%%%%%%%%%%%%%%%%%%%%%%%%%%%%%%%%%%%%%%%%%%%%%%%%%%%%%%

\section{DISCUSSION AND CONCLUSIONS}

\begin{figure}
\centering{
\includegraphics[width=15cm]{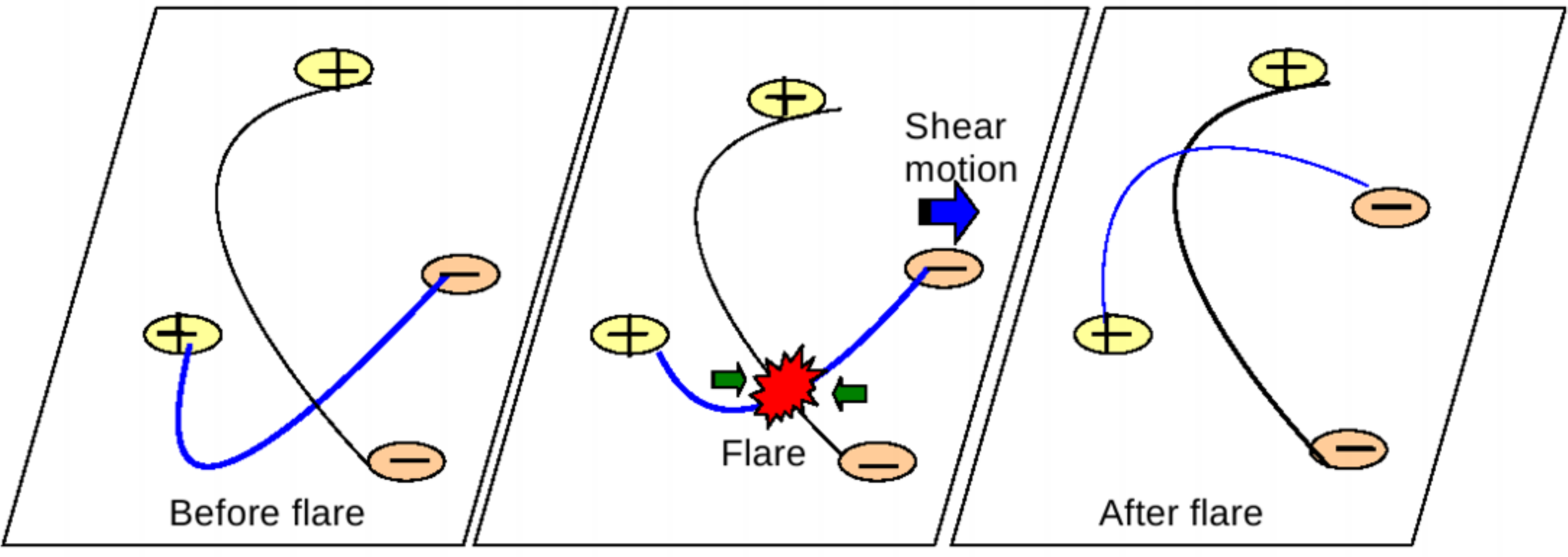}
}
\caption{Schemetic cartoons showing the flare triggering due to interaction
         of two X-ray loop-system. Black line shows the higher-loop system
         and dark blue line indicates the smaller underlying loop system.
         Due to shear motion of the right footpoint of smaller loop system,
         it becomes unstable and reconnects with the overlying higher loop
         system, triggering a flare event. After the flare event, the lower loop system becomes simplified as evident
 in GOES SXI image at 16:31:01 UT (Figure \ref{sxi}).}
\label{cartoon}
\end{figure}

We present the rare observational evidence of X-type loop-loop interaction
associated with M7.9/1N flare. %%% ? event. %%%
The coronal images observed by GOES SXI and TRACE 195~\AA \, evidently show
the interacting loop-system.
TRACE white-light images reveal the sunspots shear motion (negative
polarity) across the neutral line.
%%% across ? %%%
This shear motion probably might have produce the destabilization in the
associated loop-system and cause the loop-interaction followed by
the flare.
%%% ? event. %%%
On the basis of multi\-wavelength observations, we draw a schematic cartoon
to explain the whole event scenario (see Figure \ref{cartoon}).
%
%%% Fig. 15 %%%
%
Before the flare %%% ? event, %%%
there was two loop systems visible in SXI images.
One higher loop in N-S direction and another smaller loop system in E-W
direction lying below this higher loop system.
Due to the shear motion of the right foot\-point (anchored in negative
polarity) of smaller loop system, the loop becomes unstable and rises up
due to instability and reconnects with the overlying higher loop system
resulting X-type interaction in association with flare event.
After the flare event, the connectivity of the smaller loop system changed
into the relaxed state.

The regular variation of 4.9 and 8.8 GHz radio flux and accompanying flare
effect observed during 27 April, 2006 are interpreted using X-type loop
interaction model.
We found the oscillatory behavior with double peak structure.
Double peak in the radio flux gives the support for loop-interaction model
\citep{sakai1986}.
According to the theoretical model, the double peak structure is more
pronounced, when the currents in the two loops are sufficient for the
explosive coalescence.
Individual peak belongs to the electric field variation at the reconnection
site.
This electric field accelerates the electrons which generate the radio
emission.
The cause of quasi\-periodic oscillation is as follows: after explosive
reconnection of poloidal magnetic fields taking place at the `X' point
between approaching current loop and two plasma blobs pass through each
other and overshoot  (an approach that fails and gives way to another
attempt), resulting the repetition of the process. \citet{kliem2000} also proposed a model in which the pulsations of the radio flux are caused by quasi-periodic particle
 acceleration episodes that result from a dynamic phase of
 magnetic reconnection in a large-scale current sheet. The
 reconnection is dominated by repeated formation and sub-sequent coalescence of magnetic islands, while a continuously growing plasmoid is fed by
 newly coalescing islands. In our case, the coalescence speed of 52 km~s$^{-1}$ is much smaller than the  Alfv\'en velocity of
$\sim$1000 km~s$^{-1}$.
%the flow velocity is much below the Alfv\'en velocity in
%agreement with the observational fact that the 30 km~s$^{-1}$ colliding
%velocity of the lower loop is much smaller than the Alfv\'en velocity of
%$\sim$1000 km~s$^{-1}$.
The pre\-flare stage in which multiple current  filament structure might
be generated due to the photospheric shear motion  across the neutral
line.
The photospheric shear motion can give rise to plasma currents along the
potential magnetic field produced by the sunspots nearby the active region.
As the shear motion proceed, the current density may increase and current
loop might move up, associated with relaxation of magnetic tension \citep{sakai1986}.
The absence of type III burst during flare energy release confirms the
connectivity change and no opening of field lines.
In addition, coalescence of hard X-ray sources  also confirm the
loop-loop interaction.

\citet{sakai1986} presented the physical characteristics of the explosive
coalescence of current loops through computer simulation and theory and
mentioned canonical characteristics of the explosive coalescence as
(i) impulsive increase of kinetic energy of electrons and ions
(ii) simultaneous heating and acceleration of particles in high and low
energy spectra (i.e. Neupert effect)
(iii) quasi-periodic amplitude oscillations in field and particle quantities
(iv) a double peak (or triple peak) structure in these profiles.
Our observations clearly matches with all the above mentioned
characteristics of the explosive coalescence and provide a unique evidence
of X-type loop-loop interaction satisfying theories and simulations.

%\textit{ %%%
%
%%% Attention of the Authors !!! %%%   %%%%%%%%%%%%%%%%%%%%%%%%%%%%%%%%%%%%%%%
%
%In conclusion, we would like to draw attention of readers to a more general
%problem of solar flare energetics.
The interaction of large-scale current-carrying loops should be considered
as a part of the global electrodynamic coupling in flare-productive
active regions and active complexes as discussed in
Section~\ref{sub:topology}.
%
%%% Sub-section 2.3 %%%
%
%Why? --
On the one hand, the potential magnetic field in the corona determines a
large-scale structure of active regions while the reconnecting current
layers at separators in the corona together with other non-potential
components (see Section 14.5 in Somov, 2007) of magnetic field determine
energetic and dynamics of large flares.
On the other hand, two large-scale current-carrying loops emerging from
under the photosphere have the sufficient energy to provide a large flare
too by their interaction and coalescent instability as considered
in this paper.
Moreover, these two currents could be incorporated in the large-scale
structure with reconnecting current layer.
%
%\/}

%
%%% Attention of the Authors !!! %%%   %%%%%%%%%%%%%%%%%%%%%%%%%%%%%%%%%%%%%%%
%
The principal question is in the relative role of two distinct sources of
free magnetic energy: the interaction of magnetic fluxes, and
the interaction of electric currents as demonstrated in this paper.
Clearly the answer depends on the relation between:
(a) the photospheric flows which create the pre\-flare current layers at
the separators,
(b) the photospheric shear flows which induce the current layers extending
along the separatrices \citep{somov2002}, and
(c) the other photospheric flows like sunspot rotations which twist the
magnetic flux tubes.
In any case, the separator is a special place where a fast conversion of
free magnetic energy into bulk plasma motions, heat flows and energy of
accelerated particles can take place.

In conclusions, we find the rare multiwavelength observational signature of the loop-loop
interaction and triggering of the M-class flare, which is consistent with the earlier developed theories and simulations. However, further detailed multiwavelength studies should be carried out statistically 
by analyzing such events to shed more lights on the dynamics and energetics 
related to the flare and eruptive phenomena related to loop-loop interactions.

\acknowledgments
 We express our gratitude to the referee for his/her valuable suggestions which improved the manuscript considerably.
We acknowledge to space missions, GOES, SOHO/MDI, TRACE and RHESSI for
providing the data used in this study.
SOHO is a project of international cooperation between ESA and NASA.
We are thankful for the radio data obtained from RSTN network
(Sagamore Hill) and radiospectrograph from Ondrejov, Czech republic. We are thankful for Meudon H$\alpha$ data used in this study.  Global High Resolution H$\alpha$ Network is operated by the Space Weather Research Lab, New Jersey Institute of Technology. AKS thanks SP2RC, Department of Applied Mathematics, The University of Sheffield for
 the support of collaborative visit, where
the part of present research work has been carried out. AKS also acknowledges the joint DST-RFBR (INT/RFBR/P-38) project grant for the financial support of this work. BVS thanks the Russian Foundation for Fundamental Research (grant no. 08-02-01033). RE acknowledges M. K\'eray for patient encouragement and is also
 grateful to NSF, Hungary (OTKA, Ref. No. K67746) for financial
support received. We also thank Dr. Marc DeRosa for his valuable suggestions and discussions 
regarding the use of PFSS technique.

\bibliographystyle{apj}
\bibliography{reference1}
\clearpage

%% the file, uncomment the \plotone, \plottwo, and
%% If you need a layout that cannot be achieved with \plotone or
%% \plottwo, you can invoke the graphicx package directly with the
%% \includegraphics command or use \plotfiddle. For more information,
%% please see the tutorial on "Using Electronic Art with AASTeX" in the
%% documentation section at the AASTeX Web site,
%% http://www.journals.uchicago.edu/AAS/AASTeX.
%%
%% The examples below also include sample markup for submission of
%% supplemental electronic materials. As always, be sure to check
%% the instructions to authors for the journal you are submitting to
%% for specific submissions guidelines as they vary from

%% This example uses \plotone to include an EPS file scaled to
%% 80% of its natural size with \epsscale. Its caption
%% has been written to indicate that additional figure parts will be
%% available in the electronic journal.

%*****************************************************************************
%%%%%% Figure 1 %%%%%%   %%%%%%%%%%%%%%%%%%%%%%%%%%%%%%%%%%%%%%%%%%%%%%%%%%%%%

%*****************************************************************************
%%%%%% Figure 2 %%%%%%   %%%%%%%%%%%%%%%%%%%%%%%%%%%%%%%%%%%%%%%%%%%%%%%%%%%%%
%%%%%%%%%%%%%%%%%%%%%%%%%%%%%%%%%%%%%%%%%%%%%%%%%%%%%%%%%%%%%%%%%%%%%%%%%%%%%%

%*****************************************************************************
%%%%%% Figure 3 %%%%%%   %%%%%%%%%%%%%%%%%%%%%%%%%%%%%%%%%%%%%%%%%%%%%%%%%%%%%
%
%%%%%%%%%%%%%%%%%%%%%%%%%%%%%%%%%%%%%%%%%%%%%%%%%%%%%%%%%%%%%%%%%%%%%%%%%%%%%%

%%%%%% Figure 4 %%%%%%   %%%%%%%%%%%%%%%%%%%%%%%%%%%%%%%%%%%%%%%%%%%%%%%%%%%%%

\clearpage
%*****************************************************************************
%%%%%% Figure 5 %%%%%%   %%%%%%%%%%%%%%%%%%%%%%%%%%%%%%%%%%%%%%%%%%%%%%%%%%%%%
%

%*****************************************************************************
%%%%%% Figure 6 %%%%%%   %%%%%%%%%%%%%%%%%%%%%%%%%%%%%%%%%%%%%%%%%%%%%%%%%%%%%
%

%*****************************************************************************
%%%%%% Figure 7 %%%%%%   %%%%%%%%%%%%%%%%%%%%%%%%%%%%%%%%%%%%%%%%%%%%%%%%%%%%%
%

%*****************************************************************************
%************************************************************************************
%%%%%%%%%%%%%%%%%%%%%%%%%%%%%%%%%%%%%%%%%%%%%%%%%%%%%%%%%%%%%%%%%%%%%%%%%%%%%%
%
%*****************************************************************************

%************************************************************************************
%%%%%%%%%%%%%%%%%%%%%%%%%%%%%%%%%%%%%%%%%%%%%%%%%%%%%%%%%%%%%%%%%%%%%%%%%%%%%%
%*****************************************************************************

%*****************************************************************************
\clearpage

%*****************************************************************************
%%%%%%%%%%%%%%%%%%%%%%%%%%%%%%%%%%%%%%%%%%%%%%%%%%%%%%%%%%%%%%%%%%%%%%%%%%%%%%

%**************************************************************************
%%%%%%%%%%%%%%%%%%%%%%%%%%%%%%%%%%%%%%%%%%%%%%%%%%%%%%%%%%%%%%%%%%%%%%%%%%%

%*****************************************************************************
%%%%%%%%%%%%%%%%%%%%%%%%%%%%%%%%%%%%%%%%%%%%%%%%%%%%%%%%%%%%%%%%%%%%%%%%%%%%%%

\clearpage
%*****************************************************************************
%%% Figure 11 %%%   %%%%%%%%%%%%%%%%%%%%%%%%%%%%%%%%%%%%%%%%%%%%%%%%%%%%%%%%%%
%*****************************************************************************

%*****************************************************************************
%%% Figure 12 %%%   %%%%%%%%%%%%%%%%%%%%%%%%%%%%%%%%%%%%%%%%%%%%%%%%%%%%%%%%%%
%*****************************************************************************

%%%%%%%%%%%%%%%%%%%%%%%%%%%%%%%%%%%%%%%%%%%%%%%%%%%%%%%%%%%%%%%%%%%%%%%%%%%%%%

%*****************************************************************************
\end{document}